\documentclass[preprint]{aastex}

\usepackage[english]{babel}
\usepackage{amsmath}
\usepackage{amssymb}
\usepackage{epsfig}
\usepackage{graphicx}
\usepackage{natbib}
\usepackage{multirow}
\usepackage{longtable}
\usepackage{lscape}
\usepackage{bbm}
\usepackage{dcolumn}
\usepackage[dvips]{color}
\usepackage{rotating}
\usepackage{aas_macros}

\newcommand{\capdef}{}
\newcommand{\mycaption}[2][\capdef]{\renewcommand{\capdef}{#2}%
        \caption[#1]{{\footnotesize #2}}}
\makeatletter
\renewcommand{\fnum@table}{\textbf{\tablename~\thetable}}
\renewcommand{\fnum@figure}{\textbf{\figurename~\thefigure}}
\makeatother

\newlength{\myem}
\newcounter{mysubequation}[equation]

\newcommand{\ie}{{\it i.e.}}
\newcommand{\eg}{{\it e.g.}}

\newcommand{\cf}{{\it cf.}}

\newcommand{\etc}{{\it etc.}}
\newcommand{\eq}{Eq.}

\newcommand{\fig}{Fig.}

\newcommand{\Ref}{Ref.}

\newcommand{\Sec}{Sec.}

\newcommand{\App}{App.}

\newcommand{\Tab}{Table}

\newcommand{\equ}[1]{\eq~(\ref{equ:#1})}
\newcommand{\figu}[1]{\fig~\ref{fig:#1}}

\newcommand{\dr}{\Gamma^{\mathrm{IT}}_{a \rightarrow b}}
\newcommand{\drb}{\Gamma^{\mathrm{IT}}_{p \, \gamma \rightarrow p' \, b}}
\newcommand{\dsb}{\sigma^{\mathrm{IT}}}
\newcommand{\dcb}{\chi^{\mathrm{IT}}}

\shorttitle{Simplified models for photohadronic interactions}
\shortauthors{S.~H{\"u}mmer~et~al.}

\begin{document}
\title{
Simplified models for photohadronic interactions in cosmic accelerators
}

\author{S.~H{\"u}mmer}
\email{svenja.huemmer@physik.uni-wuerzburg.de}

\author{M.~R{\"u}ger}
\email{mlrueger@astro.uni-wuerzburg.de}

\author{F.~Spanier}
\email{fspanier@astro.uni-wuerzburg.de}

\and

\author{W.~Winter}
\email{winter@physik.uni-wuerzburg.de}
\affil{Institut f{\"u}r Theoretische Physik und Astrophysik, \\ Universit{\"a}t W{\"u}rzburg,
       97074 W{\"u}rzburg, Germany}

\begin{abstract}

We discuss simplified models for photo-meson production
in cosmic accelerators, such as Active Galactic Nuclei and Gamma-Ray Bursts. Our self-consistent models are
directly based on the underlying physics used in the SOPHIA software, and can be
easily adapted if new data are included. They allow for the efficient
computation of neutrino and photon spectra (from $\pi^0$ decays),
as a major requirement of modern time-dependent simulations of
the astrophysical sources and parameter studies. In addition, the
secondaries (pions and muons) are explicitely generated, a necessity
if cooling processes are to be included. For the neutrino production, we
include the helicity dependence of the muon decays which in fact leads to larger corrections than the details of the interaction model.
The separate computation
of the $\pi^0$, $\pi^+$, and  $\pi^-$ fluxes allows, for instance,
 for flavor ratio predictions
of the neutrinos at the source, which
are a requirement of many tests of neutrino
properties using astrophysical sources.
We confirm that for charged pion generation, the often used
production by the
$\Delta(1232)$-resonance is typically not the dominant process in Active Galactic Nuclei and Gamma-Ray Bursts,
and we show, for arbitrary input spectra, that the number of neutrinos are underestimated by at least a factor
of two if they are obtained from the neutral to charged pion ratio.
We compare our results for several levels of
simplification using isotropic synchrotron and thermal spectra, and we demonstrate that they are sufficiently
close to the SOPHIA software.
\end{abstract}


\section{Introduction}

Photohadronic interactions in high-energy astrophysical accelerators are, besides proton-proton interactions, the key ingredient of hadronic source models. The smoking gun signature of these interactions may be the neutrino production from charged pions, which could be detected in neutrino telescopes~\citep{Aslanides:1999vq, Ahrens:2002dv,Tzamarias:2003wd, Piattelli:2005hz}, such as IceCube; see, \eg, \citet{Rachen:1998fd} for the general theory of the astrophysical sources. For instance, in GRBs, photohadronic interactions are expected to lead to a significant flux of neutrinos in the fireball scenario~\citep{Waxman:1997ti}. On the other hand, in AGN models, the neutral pions produced in these interactions may describe the second hump in the observed photon spectrum, depending on the dominance of synchrotron or inverse Compton cooling of the electrons.
The protons in these models are typically assumed to be accelerated in the relativistic outflow together with electron and positrons by Fermi shock acceleration. The target photon field is typically assumed to be the synchrotron photon field of the co-accelerated electrons and positrons. Also thermal photons from broad line regions, the accretion disc, and the CMB may serve as target photon field. The latter case, relevant for the cosmogenic neutrino flux, is not considered in detail.

The basic ideas of complete hadronic models for AGN have been described already in previous works \citep{Mannheim:1993,Mucke:2000rn,Aharonian:2002} , as well as leptonic models have been discussed by \citet{Maraschi:1992,Dermer:1993}. In the first place, these models have been used as static models to describe steady-state spectral energy distributions. But with today's generation of gamma-ray telescopes, a detailed analysis of the dynamics of very high energetic sources is possible, and time-dependent modeling is inevitable. For the case of leptonic models, see, \eg, \citet{Bottcher:2002, Ruger:2010}. A necessary prerequisite for a time-dependent hadronic modeling is the efficient computation of the photohadronic interactions. The on-line calculation via Monte Carlo simulations is not feasible, therefore a parametric description is the most viable way; see, \eg, \citet{Kelner:2008ke}.

The prediction of neutrino fluxes in many source models relies on the photohadronic neutrino production.
In this case, astrophysical neutrinos are normally assumed
to originate from pion decays, with a flavor ratio at the source of
$(f_e,\, f_\mu,\, f_\tau) \simeq (1/3,\, 2/3,\, 0)$ arising from the
decays of both primary pions and secondary muons. However, it was pointed out in \Ref~\citep{Rachen:1998fd, Kashti:2005qa} that
such sources may become opaque to muons at higher energies, in which
case the flavor ratio at the source changes to $(f_e,\, f_\mu,\,
f_\tau) \simeq (0,\, 1,\, 0)$. Therefore, one
can expect a smooth transition from one type of source to the other as
a function of the neutrino energy \citep{Kachelriess:2006fi,Lipari:2007su,Kachelriess:2007tr}, depending
on the cooling processes of the intermediate muons, pions, and kaons. Recently,
the use of flavor information has been especially proposed to extract some information on the particle physics properties of the neutrinos and the properties of the source; see \Ref~\citep{Pakvasa:2008nx} for a review.
For instance, if the neutrino telescope has
some flavor identification capability, this property can be used to
extract information on the decay \citep{Beacom:2002vi, Lipari:2007su, Majumdar:2007mp,Maltoni:2008jr,Bhattacharya:2009tx}
and oscillation \citep{Farzan:2002ct, Beacom:2003zg,Serpico:2005sz, Serpico:2005bs,Bhattacharjee:2005nh, Winter:2006ce, Majumdar:2006px,Meloni:2006gv, Blum:2007ie,
Rodejohann:2006qq, Xing:2006xd, Pakvasa:2007dc, Hwang:2007na, Choubey:2008di,Esmaili:2009dz}
parameters, in a way which might be synergistic to terrestrial
measurements. Of course, the flavor ratios may be also used for source identification,
see, \eg, \citet{Xing:2006uk,Choubey:2009jq}. Except from flavor identification,
the differentiation between neutrinos and antineutrinos could be useful for the
discrimination between $p \gamma$ and $pp$ induced neutrino fluxes, or for the
test of neutrino properties (see, \eg, \citealt{Maltoni:2008jr}). A useful
observable may be the Glashow resonance process $\bar{\nu}_e + e^- \to W^- \to
\text{anything}$ at around $6.3 \, \text{PeV}$~\citep{Learned:1994wg,Anchordoqui:2004eb, Bhattacharjee:2005nh}
to distinguish between neutrinos and antineutrinos in the detector. Within the
photohadronic interactions, the $\pi^+$ to $\pi^-$ ratio determines the ratio between
electron neutrinos and antineutrinos at the source. Therefore, a useful source model for these
applications should include accurate enough flavor ratio predictions, including the possibility to include the cooling of the intermediate particles, as well as $\pi^+$ to $\pi^-$ ratio predictions. In addition,  the computation of the neutrino fluxes should be efficient enough to allow for reasonable parameter studies or to be used as a fit model.

Photohadronic interactions in astrophysical sources are typically either described by the refined Monte-Carlo simulation of the SOPHIA software~\citet{Mucke:1999yb}, which is partially based on \citet{Rachen:1996ph}, or are in very simplified approaches computed with  the $\Delta$-resonance approximation
\begin{equation}
p + \gamma \rightarrow \Delta^+ \rightarrow \left\{
\begin{array}{ll}
n + \pi^+ & 1/3 \, \, \text{of all cases} \\
p + \pi^0 & 2/3 \, \, \text{of all cases} \\
\end{array}
\right. .
\label{equ:ds}
\end{equation}
The SOPHIA software probably provides the best state-of-the-art implementation of the photo-meson production, including not only the $\Delta$-resonance, but also higher resonances, multi-pion production, and direct ($t$-channel) production of pions. Kaon production is included as well by the simulation of the corresponding QCD processes (fragmentation of color strings). The treatment in \equ{ds}, on the other hand, has been considered sufficient for many purposes, such as estimates for the neutrino fluxes. However, both of these approaches have disadvantages: The statistical Monte Carlo approach in SOPHIA is too slow for the efficient use in every step of modern time-dependent source simulations of AGNs and GRBs. The treatment in \equ{ds}, on the other hand, does not allow for predictions of the neutrino-antineutrino ratio and the shape of the secondary particle spectra, because higher resonances and other processes contribute significantly to these.
One possibility to obtain a more accurate model is to use different interaction types with different (energy-dependent) cross sections  and inelasticities (fractional energy loss of the initial nucleon). For example, one may define an interaction type for the resonances, and an interaction type for multi-pion production (see, \eg, \citet{Reynoso:2008gs,1989A&A...221..211M} and others). Typically, such approaches do not distinguish $\pi^+$ from $\pi^-$ production. An interesting alternative has been proposed in \citet{Kelner:2008ke}, which approximates the SOPHIA treatment analytically. It provides a simple and efficient way to compute the electron, photon, and neutrino spectra, by integrating out the intermediate particles. Naturally, the cooling of the intermediate particles cannot be included in such an approach.

Since we are also interested in the cooling of the intermediate particles (muons, pions, kaons), we follow a different direction. We propose a simplified model based on the very first physics principles, which means that the underlying interaction model can be easily adapted if new data are provided. In this approach, we include the intermediate particles explicitely. We illustrate the results for the neutrino spectra, but the extension to photons and electrons/positrons is straightforward.

More explicitly, the requirements for our interaction model are the following:
\begin{itemize}
\item
 The model should predict the $\pi^+$, $\pi^-$, and $\pi^0$ fluxes separately, which is needed
for the prediction of photon, neutrino, and antineutrino fluxes, and their ratios.
\item
 The model should be fast enough for time-dependent calculations and for systematic parameter studies. This, of course, requires compromises. For example, compared to SOPHIA, our kinematics treatment will be much simpler.
\item
 The particle physics properties should be transparent, easily adjustable and extendable.
\item
 The model should be applicable to arbitrary proton and photon input spectra.
\item
 The secondaries (pions, muons, kaons) should not be integrated out, because a) their synchrotron emission may contribute to observations, b) the muon (and pion/kaon) cooling affects the flavor ratios of neutrinos, and c) pion cooling may be in charge of a second spectral break in the prompt GRB neutrino spectrum, see~\cite{Waxman:1998yy}.
\item
 The cooling and escape timescales of the photohadronic interactions as well as the proton/neutron re-injection rates should be provided, since these are needed for time-dependent and steady-state models.
\item
 The kaons leading to high energy neutrinos should be roughly provided, since their different cooling properties  may lead to changes in the neutrino flavor ratios  (see, \eg, \citealt{Kachelriess:2006fi,Lipari:2007su,Kachelriess:2007tr}). Therefore, we incorporate a simplified kaon production treatment to allow for the test of the impact of such effects. This also serves as a test case for how to include new processes.
\end{itemize}
For the underlying physics, we mostly follow similar principles as in SOPHIA.

This work is organized as follows: In \Sec~\ref{sec:basics}, we review the basic principles of the particle interactions, in particular, the necessary information (response function) to compute the meson photo-production for arbitrary photon and proton input spectra. In \Sec~\ref{sec:photo}, we summarize the meson photo-production as implemented in the SOPHIA software, but we simplify the kinematics treatment. In \Sec~\ref{sec:simplemodels}, we define simplified models based on that. As a key component, we define appropriate interaction types such that we can factorize the two-dimensional response function. In \Sec~\ref{sec:comparison}, we then summarize the weak decays and compare the different approaches with each other and with the output from the SOPHIA software. For the sake of completeness, we provide in \App~\ref{app:others} how the cooling and escape timescales, and how the neutron/proton re-injection can be computed in our simplified models.
This study is supposed to be written for a broad target audience, including particle and astrophysicists.

\section{Basic principles of the particle interactions}
\label{sec:basics}

For the notation, we follow \citet{Lipari:2007su}, where we first focus on weak decays, such as $\pi^+ \rightarrow \mu^+ + \nu_\mu$, and then extend the discussion to photohadronic interactions.
Let us first consider a single decay process.
The production rate $Q_b(E_b)$ (per energy interval) of daughter particles of species $b$ and energy $E_b$ from the decay of the parent particle $a$ can be written as
\begin{equation}
Q_b(E_b) = \sum_{\mathrm{IT}} \int dE_a \, N_a(E_a) \, \dr (E_a) \, \frac{d n_{a \rightarrow b}^{\mathrm{IT}}}{dE_b} (E_a,E_b) \, .
\label{equ:prod}
\end{equation}
Here $N_a(E_a)$ is the differential spectrum of parent particles\footnote{In steady state models, this spectrum is typically obtained as the steady state spectrum including injection on the one hand side, and cooling/escape processes on the other side.} (particles per energy interval), and $\dr$ is the interaction or decay rate (probability per unit time and particle) for the process $a \rightarrow b$ as a function of energy $E_a$ (which is assumed to be zero below the threshold). Since in pion photoproduction many interaction types contribute, we split the production probability in interaction types ``IT''.

The function $d n_{a \rightarrow b}^{\mathrm{IT}}/dE_b (E_a,E_b)$ describes the distribution (as a function of parent and daughter energy) of daughter particles of type $b$ per final state energy interval $dE_b$. This function can be non-trivial. It contains the kinematics of the decay process, \ie, the energy distribution of the discussed daughter particle, other species, we are not interested in, are typically integrated out.
If more than one daughter particle of the same species $b$ is produced, or less than one (in average) because of other branchings, it must also give the number of daughter particles per event as a function of energy, which is often called ``multiplicity''.

Note that the decay can be calculated in different frames, such as the parent rest frame (PRF), in the center of mass frame of the parents (CMF), or in the shock rest frame (SRF), typically used to describe shock accelerated particles (such as from Fermi shock acceleration) in astrophysical environments. However, the cross sections, entering the interaction rate $\Gamma$, are often given in a particular frame, which has to be properly included. In addition, note that $Q$ and $N$ are typically given per volume, but here this choice is arbitrary since it enters on both sides of \equ{prod}.

Sometimes it is useful to consider the interaction or decay chain $a \rightarrow  b \rightarrow c$ without being interested in $b$, for instance, for the decay chain $\pi \rightarrow \mu \rightarrow \nu$. In this case, one can integrate out $b$.
This approach has, for instance, been used in \citet{Kelner:2008ke} for obtaining the neutrino spectra. Note, however, that the parent particles $a$ or $b$ may lose energy before they interact or decay, such as by synchrotron radiation. We do not treat such energy losses in this study explicitely, but we provide a framework to include them, such as in a steady state or time-dependent model for each particle species.

In meson photoproduction, a similar mechanism can be used. The interaction rate \equ{prod} can be interpreted in terms of the incident protons (or neutrons) because of the much higher energies in the SRF, \ie, the species $a$ is identified with the proton or neutron, which we further on abbreviate with $p$ or $p'$ (for proton or neutron).  In this case, the interaction rate depends on the interaction partner, the photon, as
\begin{equation}
\drb (E_p) = \int d \varepsilon \int\limits_{-1}^{+1} \frac{d \cos \theta_{p \gamma}}{2} \, (1- \cos \theta_{p \gamma}) \, n_\gamma(\varepsilon,\cos \theta_{p \gamma}) \, \dsb(\epsilon_r) \, .
\label{equ:pgamma}
\end{equation}
Here $n_\gamma(\varepsilon,\cos \theta_{p \gamma})$  is the photon density as a function of photon energy $\varepsilon$ and the angle between the photon and proton momenta $\theta_{p \gamma}$, $\dsb(\epsilon_r)$ is the photon production cross section, and $\epsilon_r=E_p \varepsilon/m_p (1 - \cos_{p \gamma})$ is the photon energy in the nucleon/parent rest frame (PRF) in the limit $\beta_p\approx1$. The interaction itself, and therefore $E_p$ and $\varepsilon$, is described in the SRF. The daughter particles $b$ are typically $\pi^+$, $\pi^-$, $\pi^0$, or kaons. If intermediate resonances are produced, we integrate them out so that only pions (kaons) and protons or neutrons remain as the final states.

In the following, we will assume isotropy $n_\gamma(\varepsilon,\cos \theta_{p \gamma}) \simeq n_\gamma(\varepsilon)$ of the photon distribution. This limits this specific model to scenarios where we have seed photons produced in the shock rest frame (\ie, synchrotron emission). The other interesting case, where thermal photons are coming from outside the shock, may be easily implemented as well when the shock speed is high enough to assume delta-peaked angular distributions. Arbitrary angular distributions would require additional integrations.
Assuming that the photon distribution is isotropic, the integral over $\cos \theta_{p \gamma}$ can be replaced by one over $\epsilon_r$, we have
\begin{equation}
\drb (E_p) = \frac{1}{2} \frac{m_p^2}{E_p^2} \int\limits_{\frac{\epsilon_{\mathrm{th}} m_p}{2 E_p}}^{\infty} d \varepsilon \frac{n_\gamma(\varepsilon)}{\varepsilon^2} \int\limits_{\epsilon_{\mathrm{th}}}^{2 E_p \varepsilon/m_p} d \epsilon_r \, \epsilon_r  \, \dsb(\epsilon_r) \, .
\label{equ:pgamma2}
\end{equation}
Here $\epsilon_{\mathrm{th}} \simeq 150 \, \mathrm{MeV}$ is the threshold below which the cross sections are zero.
Note that  $\epsilon_r$ corresponds to the available center of mass energy $\sqrt{s}$ of the interaction as
\begin{equation}
s(\epsilon_r) = m_p^2 + 2 \, m_p \, \epsilon_r \, .
\label{equ:s}
\end{equation}
In \equ{pgamma2}, the integral over $\epsilon_r$ takes into account that the proton and photon may hit each other in different directions. In the most optimistic case, $\epsilon_r\approx2 E_p \varepsilon/m_p$ (head-on hit), whereas in the most pessimistic case, $\epsilon_r\approx0$ (photon and proton travel in same direction). It should be noted that the assumption of isotropy here is limiting the model to internally produced photon fields. This could be lifted for different angular distributions, but except for the case of uni-directional photons this would require one more integration.

In general, the function $d n_{a \rightarrow b}^{\mathrm{IT}}/dE_b (E_a,E_b)$ in \equ{prod} is a non-trivial function of two variables $E_a$ and $E_b$. For photo-meson production in the SRF, the energy of the target photons is typically much smaller than the incident nucleon energy, and $\beta_a \simeq 1$. In this case,\footnote{In fact, the ultra-relativistic argument justifies the introduction of a
distribution function of only one final state variable
$ dn^{\mathrm{IT}}_{a\rightarrow b}/d E_b (E_a, E_b) \rightarrow M_b^{IT} (E_a) p(E_b/E_a;X_1^{IT}(E_a),\dots , X_k^{IT}(E_a))$,
where $p$ is some parameterized probability distribution function of arbitrary
shape and the parameters $X_k$ only depend on the initial state. The $\delta$-approximation is the simplest approximation, which will only be useful if the probability distribution is sufficiently peaked around its mean.
}
\begin{equation}
\frac{d n_{a \rightarrow b}^{\mathrm{IT}}}{dE_b} (E_a,E_b) \simeq \delta(E_b -  \chi_{a \rightarrow b}^{\mathrm{IT}} \, E_a ) \cdot M_b^{\mathrm{IT}}  \, .
\label{equ:nabsimple}
\end{equation}
The function $\chi_{a \rightarrow b}^{\mathrm{IT}}$, which depends on the kinematics of the process, describes  which (mean) fraction of the parent energy is deposited in the daughter species. The function  $M_b^{\mathrm{IT}}$ describes how many daughter particles are produced at this energy in average.
 For our purposes, it will typically be a constant number which depends on interaction type and species $b$ (if it changes at a certain threshold, one can define different interaction types below and above the threshold). For the relationship between $\chi_{a \rightarrow b}^{\mathrm{IT}}$ and the inelasticity $K$ (fractional energy loss of the initial nucleon), see \App~\ref{app:others}.
As we will discuss later, \equ{nabsimple} is a over-simplified for more realistic kinematics, such as in direct or multi-pion production, since $\chi_{a \rightarrow b}^{\mathrm{IT}}$ is, in general, a more complicated function of $E_b/E_a$ and the initial state properties. In this case, the distribution is not sufficiently peaked around its mean, and the $\delta$-distribution in \equ{nabsimple} has to be replaced by a broader distribution function describing the distribution of the daughter energies. Instead of choosing a broader distribution function at the expense of efficiency, we will in these case define different interaction types with different values of $\chi_{a \rightarrow b}^{\mathrm{IT}}$, simulating the broad energy distribution after the integration over the input spectra.

As the next step, we insert Eqs.~(\ref{equ:nabsimple}) and~(\ref{equ:pgamma2}), valid for photoproduction of pions, into \equ{prod}, in order to obtain:
\begin{equation}
Q_b(E_b) = \int\limits_{E_b}^{\infty} \frac{dE_p}{E_p} \, N_p(E_p) \, \int\limits_{\frac{\epsilon_{\mathrm{th}} m_p}{2 E_p}}^{\infty} d\varepsilon \, n_\gamma(\varepsilon) \,  R_b( x,y )
\label{equ:prodmaster}
\end{equation}
with
\begin{equation}
 x \equiv \frac{E_b}{E_p} \, , \quad  y\equiv \frac{E_p \, \varepsilon}{m_p} \, ,
\end{equation}
 and the ``response function''
\begin{equation}
R_b(x,y) \equiv \sum_{\mathrm{IT}} R^{\mathrm{IT}}(x,y) \equiv \sum_{\mathrm{IT}} \frac{1}{2 y^2} \int\limits_{\epsilon_{\mathrm{th}}}^{2 y} d \epsilon_r \, \epsilon_r \, \dsb(\epsilon_r) \, M^{\mathrm{IT}}_b(\epsilon_r) \, \, \delta \left( x - \dcb(\epsilon_r)  \right) \, .
\label{equ:response}
\end{equation}
Here $0 < x=E_b/E_p < 1$ is the secondary energy as a fraction of the incident proton or neutron energy, $\epsilon_{\mathrm{th}} \lesssim 2 y = 2 E_p \varepsilon/m_p \lesssim 10^{4} \, \mathrm{GeV}$ corresponds to the maximal available center of mass energy\footnote{This range is given by the range the cross sections are known. For higher energies, our model can only be used as an estimate.}, and the $\dcb$  is the fraction of proton energy deposited in the secondary. Note that $\dcb(\epsilon_r)$  and $M_b(\epsilon_r)$ are typically functions of $\epsilon_r$, if they only depend on the center of mass energy of the interaction. We will define our interaction types such that $M_b$ is independent of $\epsilon_r$.

If the response function in \equ{response} is known, the secondary spectra can be calculated from \equ{prodmaster} for {\bf arbitrary} injection and photon spectra. A similar approach was used in \citet{Kelner:2008ke} for the neutrinos and photons directly. However, we do not integrate out the intermediate particles, which in fact leads to a rather complicated function $R(x,y)$, summed over various interaction types. We show in \figu{xsec} the cross section as a function of $\epsilon_r$ for these interaction types separately, where the baryon resonances have been summed up. Naively, one would just choose the dominating contributions in the respective energy ranges in order to obtain a good approximation for $\dsb(\epsilon_r)$. However, the different contributions have different characteristics, such as different $\pi^+$ to $\pi^-$ ratios in the final states, and therefore different neutrino-antineutrino ratios. In addition, the function $\delta ( x - \dcb(\epsilon_r))$ in \equ{response} maps the same region in $\epsilon_r$ in different regions of $x=E_b/E_p$, depending  on the interaction type. This means that while a particular cross section may dominate for certain energies $\epsilon_r$, for instance, the pion energies where the interaction products are found can be very different, leading to distinctive features. Therefore, for our purposes (such as flavor ratio and neutrino-antineutrino predictions), it is not a sufficient approximation to just choose the dominating cross section.

From the particle physics point of view, $R_b(x,y)=R_b(E_b, E_p, \varepsilon)$  is very similar to a detector response function in a fixed target experiment. It describes the reconstructed particle energy spectrum as a function of the properties of the incident proton beam. As the major difference, the ``detector material'' is kept as a variable function of energy, leading to the second integral over the photon density.

\section{Review of the photohadronic interaction model}
\label{sec:photo}

 The description of photohadronic interactions is based on \citet{Rachen:1996ph,Mucke:1999yb},
 which means that the fundamental physics  is similar to SOPHIA. However, our kinematics will be somewhat simplified. The purpose of this section is to show the key features of the interaction model. In addition, it is the first step towards an analytical description for the response function in \equ{response}, which should be as simple as possible for our purposes, and as accurate as necessary.  Note that we do not distinguish between protons and neutrons for the cross sections (there are some small differences in the resonances and the multi-pion production).

\subsection{Summary of processes}

\begin{figure}[t]
\begin{center}
\includegraphics[width=0.85\textwidth]{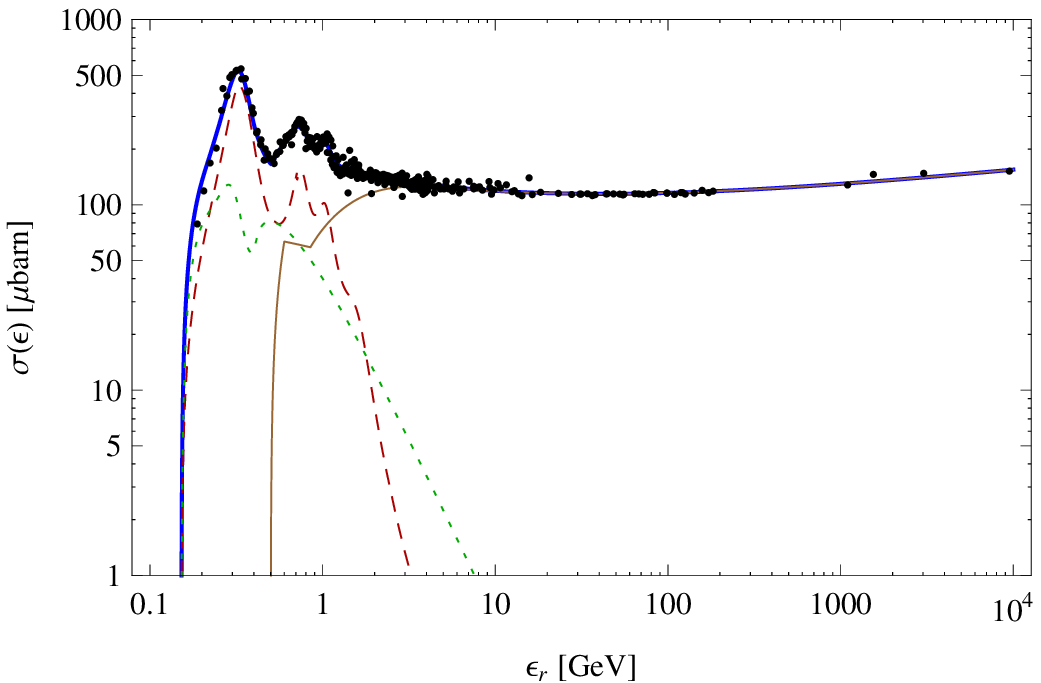}
\mycaption{\label{fig:xsec} The total $p\gamma$ photo-meson cross section
as a function of the photon's energy in the proton rest frame $\epsilon_r$
analog to \citet{Mucke:1999yb} ($1\mu$barn = $10^{-30}$ cm$^{2}$; data, shown as dots, from \citet{Amsler:2008zzb})). The contributions of baryon
resonances (red, dashed), the direct channel (green, dotted)
and multi-pion production (brown) are shown separately. }
\end{center}
\end{figure}

In summary, we consider the following processes:
\begin{description}
\item[$\boldsymbol{\Delta}$-resonance region]
The dominant resonance process is the $\Delta(1232)$-resonance (at $\epsilon_r=340 \, \text{MeV}$; \cf, \equ{s} for the $\epsilon_r$-$s$-conversion):
\begin{equation}
 p + \gamma \stackrel{\Delta(1232)}{\rightarrow} p' + \pi \, . \label{equ:delta}
\end{equation}
This process produces neutral (for $p'=p$) or charged (for $p' \neq p$) pions. For instance, for protons in the initial state, see \equ{ds}.
\item[Higher resonances]
The most important higher resonance contribution is the decay chain
\begin{equation}
\gamma +   p \stackrel{\Delta, N}{\rightarrow} \Delta' + \pi \, , \quad \Delta' \rightarrow p' + \pi' \label{equ:higherres}
\end{equation}
via $\Delta$- and $N$-resonances. Other contributions, we consider, come from the decay chain
\begin{equation}
\gamma + p \stackrel{\Delta, N}{\rightarrow} \rho + p'  \, , \quad \rho\rightarrow \pi + \pi' \, .
\label{equ:higherres2}
\end{equation}
\item[Direct production]
The same interactions as in \equ{delta} or \equ{higherres} (with the same initial and final states) can also take place in the $t$-channel, meaning that the initial $\gamma$ and nucleon exchange a pion instead of creating a (virtual) baryon resonance in the $s$-channel, which again decays.  This mechanism is also called ``direct production'', because the properties of the pion are already determined at the nucleon vertex. For instance, the process $p + \gamma\rightarrow p + \pi^0$ only takes place in the $s$-channel, whereas in the process $p+ \gamma\rightarrow n + \pi^+$ the $t$-channel is possible as well, because the photon can only couple to charged pions. The  $s$-channel reactions typically have a pronounced peak in the $\epsilon_r$-distribution, whereas they are almost structureless  in the momentum transfer distribution. On the other hand, $t$-channel reactions do not have the strong peak in the center of mass energy, but have a strong correlation between the initial and final state  momentum distributions, implying that the pions  are found at lower energies. On a logarithmic scale, however, the momentum distribution of the pions is broad.
\item[Multi-pion production]
At high energies the dominant channel is statistical multi-pion production leading to two or more pions. The process is in SOPHIA~\cite{Mucke:1999yb} described by the QCD-fragmentation of
color strings.
\end{description}
The effect of kaon decays is usually small. However, kaon decays may have interesting consequences for the neutrino flavor ratios at very high energies, in particular, if strong magnetic fields are present (the kaons are less sensitive to synchrotron losses because of the larger rest mass)~\citep{Kachelriess:2006fi,Kachelriess:2007tr}.
Therefore, we consider the leading mode: $K^+$ production (for protons in the initial state) with the decay channel leading to highest energy neutrinos.\footnote{The contributions from $K^-$ and $K^0$ are about a factor of two suppressed~\cite{Lipari:2007su}, and $K^+$ has a leading two body decay mode into neutrinos.} Note that at even higher energies, other processes, such as charmed meson production, may contribute as well.

We show in \figu{xsec} the total $p\gamma$ photo-meson cross section
as a function of $\epsilon_r$, analogously to \citet{Mucke:1999yb} ($1\mu$barn = $10^{-30}$ cm$^{2}$; data from \citet{Amsler:2008zzb})). The contributions of baryon
resonances (red, dashed), the direct channel (green, dotted)
and multi-pion production (brown) are shown separately.

In order to fully describe \equ{response} for each interaction type, we need the kinematics, entering $\chi^{\mathrm{IT}}$, the multiplicities $M_b^{\mathrm{IT}}$, and the cross section $\sigma^{\mathrm{IT}}$.

\subsection{Kinematics and secondary multiplicities}
\label{sec:kinsec}

The kinematics for the resonances and direct production can be effectively described in
the two-body picture.  We follow the calculation of \citet{Berezinsky:1993im} for the reaction $p+\gamma\rightarrow p' + \pi$ with $E_p\gg E_\gamma$ ($\beta_p\approx1$) in the SRF. The Lorentz factor between CMF and SRF is:
\begin{equation}
 \gamma_\text{cm}=\frac{E_p+E_\gamma}{\sqrt{s}}\simeq\frac{E_p}{\sqrt{s}}
\label{equ:kin1}
\end{equation}
with $\sqrt{s}$ the total CMF energy from \equ{s}. The energy of the pion in the SRF can then be written as:
\begin{equation}
 E_\pi=\gamma_\text{cm} E_\pi^\text{cm} (1+ \beta_\pi^\text{cm}\cos{\theta_\pi^\text{cm}})
= E_p \, \frac{E_\pi^\text{cm}}{\sqrt{s}}(1+\beta_\pi^\text{cm}\cos{\theta_\pi^\text{cm}})
\label{equ:kin2}
\end{equation}
with $E_\pi^\text{cm}$, $p_\pi^\text{cm}$ and $\theta_\pi^\text{cm}$ the pion energy, momentum and angle of emission in the CMF. Note that backward scattering of the pion means that $\cos{\theta_\pi^\text{cm}}=-1$.
From \equ{kin2} we can read off the  fraction of energy of the initial $p$ going into the pion $E_\pi/E_p$ (which is equal to the inelasticity of $p$ for $p'=p$).
Since $p+\gamma\rightarrow p'+\pi$ is a two body reaction, the
energies of $p'$ and pion are given by~\citet{Amsler:2008zzb}
\begin{equation}
E_{\pi}^\text{cm}=\frac{s-m_p^2+m_\pi^2}{2\sqrt{s}}
\label{equ:kin3}
\end{equation}
with $s(\epsilon_r)$ given by \equ{s}.
Using \equ{kin3} in \equ{kin2}, we can calculate $\chi$ in \equ{response} as
\begin{equation}
\label{eqn:1pion}
 \chi(\epsilon_r)=  \frac{s(\epsilon_r)-m_p^2+m_\pi^2}{2 s(\epsilon_r)} (1+\beta_\pi^{\mathrm{cm}} \cos{\theta_\pi}) \, .
\end{equation}
Indeed, $\chi(\epsilon_r)$ is a function of $\epsilon_r$ if $\cos{\theta_\pi}$ is a constant.
As described in \citet{Rachen:1996ph},  the average $\left<\cos{\theta_\pi}\right> \simeq 0$ for the resonances (such as for isotropic emission in the CMF),
whereas the direct production is backward peaked to a first approximation for sufficiently high energies $\left<\cos{\theta_\pi}\right> \rightarrow -1$. This approximation is only very crude. Therefore, we calculate the mean  $\left<\cos{\theta_\pi}\right>$ for direct production by averaging the probability distribution of the Mandelstam variable $t$, as we discuss in \App~\ref{app:directkin}.
Note that even for the resonances these scattering angle averages are only approximations; a more refined kinematics treatment, such as in SOPHIA, will lead to a smearing around these mean values.  In addition, note that we also use Eq.~(\ref{eqn:1pion}) for the kaon production (with the rough estimate $\left<\cos{\theta_\pi}\right> \simeq 0$), where the replacements $m_p\rightarrow m_\Lambda$ and $m_\pi\rightarrow m_K$ have to be made.
Similarly, one has for the first and second pion for the interaction in \equ{higherres} \citep{Rachen:1996ph}
\begin{align}
 \label{eqn:2pion1}
 \chi_a(\epsilon_r)&=\frac{s(\epsilon_r)-m_\Delta^2+m_\pi^2}{2 s(\epsilon_r)} (1+\beta_\pi^{\text{cm}} \cos{\theta_\pi}) \, ,\\
 \label{eqn:2pion2}
 \chi_b(\epsilon_r)&=\frac{s(\epsilon_r)-m_\pi^2+m_\Delta^2}{2 s(\epsilon_r)} \frac{m_\Delta^2-m_p^2+m_\pi^2}{2 m_\Delta^2} (1+\beta_\Delta^{\text{cm}} \cos{\theta_\Delta}) \, ,
\end{align}
where $\Delta = \Delta(1232)$,  and for the interaction \equ{higherres2}
\begin{align}
 \label{eqn:2pionrho}
  \chi(\epsilon_r)&=\frac{1}{2}\frac{s(\epsilon_r)-m_p^2+m_\rho^2}{2 s(\epsilon_r)} (1+\beta_\rho^{\text{cm}} \cos{\theta_\rho}) = \frac{K}{2}
\end{align}
 with $m_\rho \simeq 775 \, \mathrm{MeV}$.
Note that in order to evaluate the $\delta$-function in \equ{response} (if one integrates over $\epsilon_r$), one also needs the derivative $\chi'(\epsilon_r)$, which can easily be computed from the functions above.

\begin{table}[t]
\begin{center}
\small{
 \begin{tabular}{c||ccc|cc||ccc|cc||cc}
 \cline{2-13}
   & \multicolumn{5}{c||}{Initial $p$ is proton} & \multicolumn{5}{c||}{Initial $p$ is neutron}  & \multicolumn{2}{c}{Kinematics} \\
\hline
 IT & $M_{\pi^0}$ & $M_{\pi^+}$ & $M_{\pi^-}$ & $M_{n}$ & $M_{p}$ & $M_{\pi^0}$ & $M_{\pi^+}$ & $M_{\pi^-}$ & $M_{n}$ & $M_{p}$ & Eq. & $\left<\cos\theta\right>$ \\
\hline
\multicolumn{13}{c}{Interactions $\gamma + p \stackrel{\mathrm{IT}}{\rightarrow} p' + \pi$}  \\
\hline
  $\Delta_1$ & $\frac{2}{3}$ & $\frac{1}{3}$ & - & $\frac{1}{3}$ & $\frac{2}{3}$ &$\frac{2}{3}$ & - & $\frac{1}{3}$ &  $\frac{2}{3}$ & $\frac{1}{3}$ & (\ref{eqn:1pion}) & $0$ \\
  $N_1$ & $\frac{1}{3}$ & $\frac{2}{3}$ & - & $\frac{2}{3}$ & $\frac{1}{3}$ & $\frac{1}{3}$ & - & $\frac{2}{3}$ & $\frac{1}{3}$ & $\frac{2}{3}$ & (\ref{eqn:1pion}) & $0$ \\
 T$_1$ & - & $1$ & - & $1$ & - & - & - & $1$ & - & $1$ & (\ref{eqn:1pion})& $-1$ \\
\hline
\multicolumn{13}{c}{Interactions $\gamma + p \stackrel{\mathrm{IT}}{\rightarrow} \Delta' + \pi$, $\Delta' \rightarrow p' + \pi'$}\\
\hline
& \multicolumn{12}{c}{Properties of first pion $\pi$ and nucleon $p'$ }\\
\hline
 $\Delta_{2 \text{a}}$ & $\frac{1}{15}$ & $\frac{8}{15}$ & $\frac{2}{5}$ & $\frac{17}{45}$ & $\frac{28}{45}$ & $\frac{1}{15}$ & $\frac{2}{5}$ & $\frac{8}{15}$ & $\frac{28}{45}$ &  $\frac{17}{45}$ & (\ref{eqn:2pion1}) & $0$ \\
 $N_{2 \text{a}}$ & $\frac{1}{3}$ & $\frac{1}{6}$ & $\frac{1}{2}$ &  $\frac{2}{9}$ & $\frac{7}{9}$ & $\frac{1}{3}$ & $\frac{1}{2}$ & $\frac{1}{6}$ &  $\frac{7}{9}$ & $\frac{2}{9}$ & (\ref{eqn:2pion1}) & $0$ \\
  T$_{2 \text{a}}$ &  - & $\frac{1}{4}$ & $\frac{3}{4}$ &  $\frac{1}{6}$ & $\frac{5}{6}$ & - & $\frac{3}{4}$ & $\frac{1}{4}$ &  $\frac{5}{6}$ & $\frac{1}{6}$ & (\ref{eqn:2pion1}) & $-1$ \\
\hline
& \multicolumn{12}{c}{Properties of second pion $\pi'$}\\
\hline
  $\Delta_{2 \text{b}}$ & $\frac{2}{5}$ & $\frac{19}{45}$ & $\frac{8}{45}$ & & & $\frac{2}{5}$ & $\frac{8}{45}$ & $\frac{19}{45}$ &  & & (\ref{eqn:2pion2})&$0$ \\
  $N_{2 \text{b}}$ & $\frac{1}{3}$ & $\frac{11}{18}$ & $\frac{1}{18}$ & & & $\frac{1}{3}$ & $\frac{1}{18}$ & $\frac{11}{18}$ & & & (\ref{eqn:2pion2})& $0$ \\
  T$_{2 \text{b}}$ &  $\frac{1}{6}$ & $\frac{3}{4}$ & $\frac{1}{12}$ & & & $\frac{1}{6}$ & $\frac{1}{12}$ & $\frac{3}{4}$ & &  & (\ref{eqn:2pion2})& $1$ \\
\hline
 \multicolumn{13}{c}{Interactions $\gamma + p \stackrel{\mathrm{IT}}{\rightarrow} \rho + p' $, $\rho\rightarrow \pi + \pi'$ ($\pi$ and $\pi'$ summed over)}\\
\hline
 $\Delta_3$ & $\frac{1}{3}$ & $1$ & $\frac{2}{3}$ & $\frac{1}{3}$ & $\frac{2}{3}$ & $\frac{1}{3}$ & $\frac{2}{3}$ & $1$ & $\frac{2}{3}$ & $\frac{1}{3}$ &  (\ref{eqn:2pionrho})&$0$ \\
  $N_3$ & $\frac{2}{3}$ & $1$ & $\frac{1}{3}$ & $\frac{2}{3}$ & $\frac{1}{3}$ & $\frac{2}{3}$ & $\frac{1}{3}$ & $1$ & $\frac{1}{3}$ & $\frac{2}{3}$ &  (\ref{eqn:2pionrho})& $0$ \\
\hline
 \end{tabular}
} 
\end{center}
\mycaption{\label{tab:numbers} Summary of the average multiplicities for different types of considered resonances and direct production channels. The multiplicities of the pions $M_\pi$ add up to the number of pions produced (one or two), the multiplicities for the nucleons $M_p'$ in the final state $p'$ to one. In the last columns, the cosine of the approximate scattering angle in the center of mass frame is given, together with the equation for kinematics. The interaction types IT are labeled by the type of resonance ($\Delta$ or $N$) or ``Direct'' for the direct production ($t$-channel).
}
\end{table}

The average multiplicities for different types of resonances and direct production channels can be found in \Tab~\ref{tab:numbers}. The multiplicities of the pions $M_\pi$ add up to the number of pions produced (one or two), the multiplicities for the nucleons $M_p'$ in the final state $p'$ to one. They are needed in \equ{nabsimple} and \equ{cool} (cooling and escape timescale in \App~\ref{app:others}). In the last columns, the cosine of the approximate scattering angle in the center of mass frame is given, together with the equation for kinematics. The interaction types IT are labeled by the type of resonance ($\Delta$ or $N$) or $T$ for the direct production ($t$-channel). Note that different resonances will contribute with a similar pattern, such as through the interaction types $\Delta_1(1232)$, $\Delta_1(1700)$, $\Delta_1(1905)$, $\Delta_1(1950)$, \etc. In addition note that the branching ratios into certain resonances and from a resonance into the described channels are absorbed in the cross sections. Therefore, the total yield is always one (or two) pions per interaction in \Tab~\ref{tab:numbers}.

For the multi-pion channel, we use two different approaches. A very simple treatment can be performed similar to \citet{Atoyan:2002gu}, with some elements of \citet{Mucke:1999yb}. Most of the energy lost by the proton ($\simeq 0.6 \, E_p$, or $K \simeq 0.6$) is split equally among three pions. The three types $\pi^+$, $\pi^-$, and $\pi^0$ are therefore approximately produced in equal numbers.  Our multi-pion channel, parameterized by the multi-pion cross section in \figu{xsec}, however, is actually a combination of different processes and residual cross sections. For instance, diffractive scattering is a small contribution. Therefore, in order to reproduce the pion multiplicities times cross sections in Figs.~9 and~10 of \citet{Mucke:1999yb} more accurately, we choose $M_{\pi^0}=1$ ($M_{\pi^0}=1$), $M_{\pi^+}=1.2$ ($M_{\pi^+}=0.85$), and $M_{\pi^-} = 0.85$ ($M_{\pi^-} =1.2$) for initial protons (neutrons). Close to the threshold,
the decreasing phase space for pion production requires a modification of the threshold $\epsilon_r \ge 1 \, \mathrm{GeV}$ for $\pi^-$ ($\pi^+$) from initial protons (neutrons). This corresponds to a vanishing multiplicity below the threshold, \ie, we assume that below $\epsilon_r \ge 1 \, \mathrm{GeV}$ only two pions are produced.
We assume for the fraction of the proton energy going into one pion produced in the multi-pion channel $\chi^{\mathrm{Multi-\pi}} \simeq 0.2$.
 In addition, we choose $M_p=0.69$ and $M_n=0.31$ for initial protons (or $M_p=0.31$ and $M_n=0.69$ for initial neutrons) in accordance with Fig.~11 of \citet{Mucke:1999yb}  for high energies. As alternative, we use a more accurate but computationally more expensive approach, which is  directly based on the kinematics of the fragmentation code used by SOPHIA (\cf, \Sec~\ref{sec:mp2}).

\subsection{Cross sections}
\label{sec:xsec}

We parameterize the cross sections of photohadronic interactions following \citet{Mucke:1999yb}. We split the processes into three parts: resonant, direct, and multi-pion production. The different contributions are shown in \figu{xsec}.

The low energy part of the total cross section is dominated by excitations and decays of baryon resonances $N$ and $\Delta$. The cross sections for these processes can be described by the Breit-Wigner formula
\begin{equation}
\label{equ:breitwig}
 \sigma_{\mathrm{BW}}^{\mathrm{IT}}(\epsilon_r) =\frac{s}{(s-m_p^2)^2}\frac{4\pi (2J+1)B_\gamma B_{\mathrm{out}}s\Gamma^2}{(s-M^2)^2+s\Gamma^2}=B_{\mathrm{out}}^{\mathrm{IT}} \frac{s }{\epsilon_r^2}\frac{\sigma_0^{\mathrm{IT}}(\Gamma^{\mathrm{IT}})^2 s}{(s-(M^{\mathrm{IT}})^2)^2+(\Gamma^{\mathrm{IT}})^2 s}
\end{equation}
with $s(\epsilon_r)$ given by \equ{s}.
Here $J$, $M$, and $\Gamma$ being the spin, the nominal mass, and the width of the resonance, respectively.
We consider the energy-dependent branching ratios $B_{\mathrm{out}}$ and the resonances shown in the Appendix~B of \citet{Mucke:1999yb}; the total branching ratios are also listed in \Tab~\ref{tab:resonances}. Note that the energy-dependent branching ratios respect the different energy thresholds for different channels (\eg, interaction types~1 to~4). For simplicity, we neglect the slight cross section differences between $n\gamma$- and $p\gamma$-interactions and use the values for protons, which implies that we do not take into account the N(1675) resonance.\footnote{For a more accurate treatment for neutrons, also include the N(1675) resonance, whereas N(1650) and N(1680) do not apply. In addition, in \equ{mp1}, replace $80.3 \rightarrow 60.2$ and in \equ{mp2}, replace $29.3 \rightarrow 26.4$.} In \Tab~\ref{tab:resonances} we list all considered resonances and their contributions to the different interaction types. To account for the phase-space reduction near threshold the function $Z^{\mathrm{IT}}$ is introduced and multiplied on \equ{breitwig}:
\begin{equation}
 \label{equ:quenchf}
Z^{\mathrm{IT}}(\epsilon_r,\epsilon_{\mathrm{th}}^{\mathrm{IT}},w^{\mathrm{IT}})=\begin{cases} 0 & \text{if $\epsilon_r\leq\epsilon_{\mathrm{th}}^{\mathrm{IT}}$,}
\\
\frac{\epsilon_r-\epsilon_{\mathrm{th}}^{\mathrm{IT}}}{w^{\mathrm{IT}}} &\text{if $\epsilon_{\mathrm{th}}^{\mathrm{IT}}<\epsilon_r < w^{\mathrm{IT}}+\epsilon_{\mathrm{th}}^{\mathrm{IT}}$,}\\
1 & \text{if $\epsilon_r \ge w^{\mathrm{IT}}+\epsilon_{\mathrm{th}}^{\mathrm{IT}}$.}
\end{cases}
\end{equation}
with $\epsilon_{\mathrm{th}}^{\mathrm{IT}}$ and $w^{\mathrm{IT}}$ taking the values shown in \Tab~\ref{tab:resonances}.

\begin{table}[t]
\begin{center}
\small{
 \begin{tabular}{c||rrr||rr||ccc}
\cline{2-9}
   & \multicolumn{3}{c||}{Properties} & \multicolumn{2}{c||}{$Z$} & \multicolumn{3}{c}{Total $B_\mathrm{out}$} \\
\hline
 Resonance & $M$ [GeV] & $\Gamma$ [GeV] & $\sigma_0$ [$\mu$barn] & $\epsilon_{\mathrm{th}}$ [GeV] & $w$ [GeV] & IT $1$ & IT $2$ & IT $3$\\
\hline
  $\Delta_{\mathrm{IT}}(1232)$ & $1.231$ & $0.11$ &  $31.125$ &  $0.152$ & $0.17$ & 100\% & - & -    \\
  $N_{\mathrm{IT}}(1440)$ & $1.440$ & $0.35$  & $1.389$  & $0.152$ & $0.38$ & 67\% & 33\% & -     \\
  $N_{\mathrm{IT}}(1520)$ & $1.515$ & $0.11$  & $25.567$  & $0.152$ & $0.38$ & 52\% & 42\% & 6\%     \\
  $N_{\mathrm{IT}}(1535)$ & $1.525$ & $0.10$  & $6.948$  & $0.152$ & $0.38$ & 45\% & - & -   \\
  $N_{\mathrm{IT}}(1650)$ & $1.675$ & $0.16$  & $2.779$  & $0.152$ & $0.38$ & 75\% & 14\% & -     \\
  $N_{\mathrm{IT}}(1680)$ & $1.680$ & $0.125$  & $17.508$  & $0.152$ & $0.38$ & 64\% & 22\% & 14\%    \\
  $\Delta_{\mathrm{IT}}(1700)$ & $1.690$ & $0.29$ & $11.116$  & $0.152$ & $0.38$ & 14\% & 55\% & 31\%     \\
  $\Delta_{\mathrm{IT}}(1905)$ & $1.895$ & $0.35$  & $1.667$  & $0.152$ & $0.38$ & 14\% & 13\% & 73\%     \\
  $\Delta_{\mathrm{IT}}(1950)$ & $1.950$ & $0.30$  & $11.116$  & $0.152$ & $0.38$ & 37\% & 39\% & 24\%    \\
\hline
 \end{tabular}
} 
\end{center}
\mycaption{\label{tab:resonances} Summary of the considered resonances. Properties of the resonances and parameters for the quenching function \equ{quenchf} taken from \citet{Mucke:1999yb}.  The four rightmost columns refer to possible interaction types IT in \Tab~\ref{tab:numbers} and show the
total branching ratios. In fact, at the end, each combination of a specific resonance and a particular decay chain corresponds to a separate interaction type, such as $\Delta_{1}(1232)$. }
\end{table}
For the direct and the multi-pion cross section we adopt the formulae from SOPHIA. The direct part is given by
\begin{align}
 \sigma^{\mathrm{T}_1}(\epsilon_r)&=92.7 \, \mathrm{Pl}(\epsilon_r,0.152,0.25,2)+40\exp \left( -\frac{(\epsilon_r-0.29)^2}{0.002} \right)-15\exp \left( -\frac{(\epsilon_r-0.37)^2}{0.002} \right) \label{equ:smp1}\\
\sigma^{\mathrm{T}_2}(\epsilon_r)&=37.7 \, \mathrm{Pl}(\epsilon_r,0.4,0.6,2) \label{equ:smp2}
\end{align}
with
\begin{equation}
 \mathrm{Pl}(\epsilon_r,\epsilon_{\mathrm{th}},\epsilon_{max},\alpha)=\begin{cases}
                                                     0 & \text{if $\epsilon_r\leq\epsilon_{\mathrm{th}}$}\\
\left(\frac{\epsilon_r-\epsilon_{\mathrm{th}}}{\epsilon_{max}-\epsilon_{\mathrm{th}}}\right)^{A-\alpha}\left(\frac{\epsilon_r}{\epsilon_{max}}\right)^{-A} & \text{else}
                                                    \end{cases}
\end{equation}
where $A=\alpha\epsilon_{max}/\epsilon_{\mathrm{th}}$. The  cross sections are given  in $\mu$barns, and the interaction types are taken from \Tab~\ref{tab:numbers}, \ie, T$_1$ is the single pion direct production, and T$_2$ is the two pion direct production. The multi-pion cross section can be parameterized by the sum of the following two formulas (cross sections in $\mu$barns):
\begin{align}
 \sigma^{\mathrm{Multi-\pi_1}}(\epsilon_r)&=
80.3 \, Q_f(\epsilon_r,0.5,0.1) \, s^{-0.34} \label{equ:mp1}
 \\
\sigma^{\mathrm{Multi-\pi_2}}(\epsilon_r)&=\left[1-\exp\left(-\frac{\epsilon_r-0.85}{0.69}\right)\right]\left(29.3\, s^{-0.34}+59.3 \, s^{0.095}\right)\ \  \text{if $\epsilon_r>0.85$.}
\label{equ:mp2}
\end{align}

We have compared our resonant and direct production multiplicities times cross sections and proton to neutron ratios with SOPHIA (\cf, Figs.~9 to~11 in \citet{Mucke:1999yb}), and they are in excellent agreement.

\section{Simplified models}
\label{sec:simplemodels}

Here we discuss simplified models based on \Sec~\ref{sec:photo}, where we demand the features given in the introduction. For the resonances, we propose two methods, one is supposed to be more accurate, the other one faster.

\subsection{Factorized response function}
\label{sec:rsimple}

First of all, it turns out to be useful to simplify the response function in \equ{response}.
In our simplified approaches, we follow the following general idea: In \equ{prodmaster} and \equ{response}, in principle, (at least) two integrals have to be evaluated. Let us now split up the interactions in interaction types such that  $\chi^{\mathrm{IT}}(\epsilon_r) \equiv \chi^{\mathrm{IT}}$  and $M_b^{\mathrm{IT}} (\epsilon_r) \equiv M_b^{\mathrm{IT}}$ are approximately constants independent of $\epsilon_r$, but dependent on the interaction type. The response function in \equ{response} then factorizes in
\begin{equation}
R^{\mathrm{IT}}(x,y)= \delta(x - \chi^{\mathrm{IT}}) \,  M_b^{\mathrm{IT}} \, f^{\mathrm{IT}}(y)  \quad \text{with} \quad
  f^{\mathrm{IT}}(y) \equiv \frac{1}{2y^2} \int\limits_{\epsilon_{\mathrm{th}}}^{2y} d \epsilon_r \, \epsilon_r \, \sigma^{\mathrm{IT}}(\epsilon_r)  \, .
\label{equ:split}
\end{equation}
The part $\delta(x - \chi^{\mathrm{IT}})$  describes at what energy the secondary particle is found, whereas the part $M_b^{\mathrm{IT}} \, f^{\mathrm{IT}}(y)$ describes the production rate of the specific species $b$ as a function of $y$. If only the number of produced secondary particles is important, it is often useful to show the effective $F_b(y) \equiv \sum_{\mathrm{IT}}  M_b^{\mathrm{IT}} \, f^{\mathrm{IT}}(y)$.

As the next steps, we evaluate \equ{prodmaster} with \equ{split} by integrating over $dx/x=-dE_p/E_p$ and by re-writing the $\varepsilon$-integral as $y=E_p \varepsilon/m_p$-integral:
\begin{equation}
Q_b^{\mathrm{IT}} = N_p \left( \frac{E_b}{\chi^{\mathrm{IT}}} \right) \frac{m_p}{E_b} \int\limits_{\epsilon_{\mathrm{th}}/2}^\infty dy \, n_\gamma \left( \frac{m_p \, y \, \chi^{\mathrm{IT}}}{E_b} \right) \,  M_b^{\mathrm{IT}}  \, f^{\mathrm{IT}}(y) \, .
\label{equ:prodsimple}
\end{equation}
This (single) integral is relatively simple and fast to compute if the simplified response function $f^{\mathrm{IT}}(y)$ together with $\chi^{\mathrm{IT}}$ is given.  Therefore, the original double integral simplifies in this single integral, summed over a number of appropriate interaction types.
In the following, we therefore define $M_b^{\mathrm{IT}}$, $f^{\mathrm{IT}}(y)$ and $\chi^{\mathrm{IT}}$ for suitable interaction types.

\subsection{Resonances}

Here we describe two alternatives to include the resonances. Model~A is more accurate but slower to evaluate, because it includes more interaction types, whereas model~B is faster for computations, since there are only two interaction types defined for the resonances.

\subsubsection{Simplified model A (Sim-A)}
\label{sec:sima}

\begin{table}[t!]
\begin{center}
\begin{tabular}{lrrrr|ccc|ccc}
\hline
 & & & & & \multicolumn{3}{c|}{Initial proton} & \multicolumn{3}{c}{Initial neutron} \\
IT & $B^{\mathrm{IT}}_{\mathrm{out}}$ & $\epsilon_{r}^{\mathrm{IT}, 0}$ & $\hat{\Gamma}^{\mathrm{IT}}$& $\chi^{\mathrm{IT}}$ & $M_{\pi^0}$ & $M_{\pi^+}$ & $M_{\pi^-}$ & $M_{\pi^0}$ & $M_{\pi^+}$ & $M_{\pi^-}$ \\
\hline
 $\Delta_1$(1232) & 1.00 & 0.34 & 66.69 & 0.22 &
$\frac{2}{3}$ & $\frac{1}{3}$ & - & $\frac{2}{3}$ & - & $\frac{1}{3}$
   \\
 $N_1$(1440) & 0.67 & 0.64 & 4.26 & 0.29 &
 $  \frac{1}{3}$ & $\frac{2}{3}$ & - & $\frac{1}{3}$ & - & $\frac{2}{3}$
   \\
 $N_{2\mathrm{a}}$(1440) & 0.33 & 0.64 & 4.26 & 0.14 &
 $  \frac{1}{3}$ & $\frac{1}{6}$ & $\frac{1}{2}$ &
   $\frac{1}{3}$ & $\frac{1}{2}$ & $\frac{1}{6} $\\
 $N_{2\mathrm{b}}$(1440) & 0.33 & 0.64 & 4.26 & 0.19 &
  $ \frac{1}{3}$ & $\frac{11}{18}$ & $\frac{1}{18}$ &
  $\frac{1}{3}$ & $\frac{1}{18}$ & $\frac{11}{18} $\\
 $N_{1}$(1520) & 0.52 & 0.75 & 27.01 & 0.31 &
  $ \frac{1}{3}$ & $\frac{2}{3}$ & - & $\frac{1}{3}$ & - & $\frac{2}{3}$
   \\
 $N_{2\mathrm{a}}$(1520) & 0.42 & 0.75 & 27.01 & 0.17 &
  $\frac{1}{3}$ & $\frac{1}{6}$ & $\frac{1}{2}$ &
   $\frac{1}{3}$ & $\frac{1}{2}$ & $\frac{1}{6}$ \\
 $N_{2\mathrm{b}}$(1520) & 0.42 & 0.75 & 27.01 & 0.18 &
   $\frac{1}{3}$ & $\frac{11}{18}$ & $\frac{1}{18}$ &
   $\frac{1}{3}$ & $\frac{1}{18}$ & $\frac{11}{18}$ \\
 $N_{3}$(1520) & 0.06 & 0.75 & 27.01 & 0.22 &
   $\frac{2}{3}$ & $1$ & $\frac{1}{3}$ & $\frac{2}{3}$ & $\frac{1}{3}$ & $1$
   \\
 $N_{1}$(1535) & 0.45 & 0.77 & 6.59 & 0.32 &
   $\frac{1}{3}$ & $\frac{2}{3}$ & - & $\frac{1}{3}$ & - & $\frac{2}{3}$
   \\
 $N_{1}$(1650) & 0.75 & 1.03 & 3.19 & 0.35 &
   $\frac{1}{3}$ & $\frac{2}{3}$ & - &$ \frac{1}{3}$ & - & $\frac{2}{3}$
   \\
 $N_{2\mathrm{a}}$(1650) & 0.14 & 1.03 & 3.19 & 0.23 &
  $ \frac{1}{3}$ & $\frac{1}{6}$ & $\frac{1}{2}$ &
   $\frac{1}{3}$ & $\frac{1}{2}$ & $\frac{1}{6}$\\
 $N_{2\mathrm{b}}$(1650) & 0.14 & 1.03 & 3.19 & 0.17 &
  $ \frac{1}{3}$ & $\frac{11}{18}$ & $\frac{1}{18}$ &
  $ \frac{1}{3}$ & $\frac{1}{18}$ & $\frac{11}{18} $\\
 $N_{1}$(1680) & 0.64 & 1.04 & 15.72 & 0.35 &
  $ \frac{1}{3}$ & $\frac{2}{3}$ & - & $\frac{1}{3}$ & - & $\frac{2}{3}$
   \\
 $N_{2\mathrm{a}}$(1680) & 0.22 & 1.04 & 15.72 & 0.24 &
   $\frac{1}{3}$ & $\frac{1}{6}$ & $\frac{1}{2}$ &
   $\frac{1}{3}$ & $\frac{1}{2}$ & $\frac{1}{6}$ \\
 $N_{2\mathrm{b}}$(1680) & 0.22 & 1.04 & 15.72 & 0.17 &
   $\frac{1}{3}$ & $\frac{11}{18}$ & $\frac{1}{18}$ &
  $ \frac{1}{3}$ & $\frac{1}{18}$ & $\frac{11}{18} $\\
 $N_{3}$(1680) & 0.14 & 1.04 & 15.72 & 0.23 &
   $\frac{2}{3}$ & $1$ & $\frac{1}{3}$ & $\frac{2}{3}$ & $\frac{1}{3}$ & $1$
   \\
 $\Delta_{1}$(1700) & 0.14 & 1.05 & 21.68 & 0.35 &
  $ \frac{2}{3}$ & $\frac{1}{3}$ & - & $\frac{2}{3}$ & - & $\frac{1}{3}$
   \\
 $\Delta_{2\mathrm{a}}$(1700) & 0.55 & 1.05 & 21.68 & 0.24 &
  $\frac{1}{15}$ & $\frac{8}{15}$ & $\frac{2}{5}$ &
   $\frac{1}{15}$ & $\frac{2}{5}$ & $\frac{8}{15}$ \\
 $\Delta_{2\mathrm{b}}$(1700) & 0.55 & 1.05 & 21.68 & 0.16 &
   $\frac{2}{5}$ & $\frac{19}{45}$ & $\frac{8}{45}$ &
  $ \frac{2}{5}$ & $\frac{8}{45}$ & $\frac{19}{45} $\\
 $\Delta_{3}$(1700) & 0.31 & 1.05 & 21.68 & 0.23 &
  $ \frac{1}{3}$ & $1$ & $\frac{2}{3}$ & $\frac{1}{3}$ & $\frac{2}{3}$ & $1$
   \\
 $\Delta_{1}$(1905) & 0.14 & 1.45 & 3.03 & 0.38 &
  $ \frac{2}{3}$ & $\frac{1}{3}$ & - & $\frac{2}{3}$ & - & $\frac{1}{3}$
   \\
 $\Delta_{2\mathrm{a}}$(1905) & 0.13 & 1.45 & 3.03 & 0.29 &
   $\frac{1}{15}$ & $\frac{8}{15}$ & $\frac{2}{5}$ &
   $\frac{1}{15}$ & $\frac{2}{5}$ & $\frac{8}{15} $\\
 $\Delta_{2\mathrm{b}}$(1905) & 0.13 & 1.45 & 3.03 & 0.15 &
   $\frac{2}{5}$ & $\frac{19}{45}$ & $\frac{8}{45}$ &
   $\frac{2}{5}$ & $\frac{8}{45}$ & $\frac{19}{45}$ \\
 $\Delta_{3}$(1905) & 0.73 & 1.45 & 3.03 & 0.23 &
  $\frac{1}{3}$ & $1$ & $\frac{2}{3}$ & $\frac{1}{3}$ & $\frac{2}{3}$ & $1$
   \\
 $\Delta_{1}$(1950) & 0.37 & 1.56 & 16.46 & 0.39 &
  $ \frac{2}{3}$ & $\frac{1}{3}$ & - & $\frac{2}{3}$ & - & $\frac{1}{3}$
   \\
 $\Delta_{2\mathrm{a}}$(1950) & 0.39 & 1.56 & 16.46 & 0.30 &
  $ \frac{1}{15}$ & $\frac{8}{15}$ & $\frac{2}{5}$ &
  $ \frac{1}{15}$ & $\frac{2}{5}$ & $\frac{8}{15}$\\
 $\Delta_{2\mathrm{b}}$(1950) & 0.39 & 1.56 & 16.46 & 0.15 &
   $\frac{2}{5}$ & $\frac{19}{45}$ & $\frac{8}{45}$ &
  $ \frac{2}{5}$ & $\frac{8}{45}$ & $\frac{19}{45}$ \\
 $\Delta_{3}$(1950) & 0.24 & 1.56 & 16.46 & 0.23 &
   $\frac{1}{3}$ & $1$ & $\frac{2}{3}$ & $\frac{1}{3}$ & $\frac{2}{3}$ & $1$ \\
\hline
\end{tabular}
\end{center}
\mycaption{\label{tab:simmodela} Parameters for the resonant pion production in our simplified treatment~A. The units of $\epsilon_{r}^{\mathrm{IT}, 0}$ are GeV, the units of $\hat{\Gamma}^{\mathrm{IT}}$ are $\mu$barn~GeV.
}
\end{table}

In this simplified resonance treatment we make use of the fact that the resonances have cross sections peaked in $\epsilon_r$. In principle, these can be approximated from \equ{breitwig} and \equ{quenchf} by a $\delta$-function
\begin{equation}
 \sigma^{\mathrm{IT}}(\epsilon_r) \simeq   B^{\mathrm{IT}}_{\mathrm{out}} \, \hat\Gamma^{\mathrm{IT}} \,  \delta(\epsilon_r-\epsilon_{r}^{\mathrm{IT}, 0}) \quad \text{with} \quad  \hat\Gamma^{\mathrm{IT}} = \int\limits_0^{\infty} \sigma^{\mathrm{IT}}_{\mathrm{BW}} \,  Q_f^{\mathrm{IT}}(\epsilon_r,\epsilon_{\mathrm{th}}^{\mathrm{IT}},w^{\mathrm{IT}}) d \epsilon_r \, ,
\label{equ:sigmadelta}
\end{equation}
 where $\hat\Gamma^{\mathrm{IT}}$ and $\epsilon_{r}^{\mathrm{IT}, 0}$ correspond to surface area (in $\mu$barn-$\mathrm{GeV}$) and position (in GeV) of the resonance, and are process-dependent constants.
Therefore, using \equ{sigmadelta}, the $\epsilon_r$-integral in \equ{response} can be easily performed, and we obtain again the simplified \equ{split} with
\begin{equation}
f^{\mathrm{IT}}(y) = \frac{1}{2 y^2} \, \epsilon_{r}^{\mathrm{IT}, 0} \,  B^{\mathrm{IT}}_{\mathrm{out}} \, \hat\Gamma^{\mathrm{IT}}    \, \Theta(2 y -  \epsilon_r^{\mathrm{IT},0}) \quad \text{and} \quad  \chi^{\mathrm{IT}} \equiv \chi^{\mathrm{IT}}(\epsilon_r^{\mathrm{IT}, 0})\, .
\label{equ:fa}
\end{equation}
The relevant parameters for all interaction types can be read off from \Tab~\ref{tab:simmodela}.
 Note that $M_\pi^{\mathrm{IT}}$ are the total (\ie, energy-independent) branching ratios from \Tab~\ref{tab:resonances}. In addition, note that in some cases, the resonance peak may be below threshold for some of the interaction types. However, for reasonably broad energy distributions to be folded with and the total branching ratios used, this should be a good approximation. The pion spectra are finally obtained from \equ{prodsimple}, summing over all interactions listed in \Tab~\ref{tab:simmodela}.

\subsubsection{Simplified model B (Sim-B)}
\label{sec:simb}

In this model, we take into account the width of the resonances and approximate them by constant cross sections within certain energy ranges. Compared to approaches such as in \citet{Reynoso:2008gs}, we distinguish between $\pi^+$ and $\pi^-$ production (\ie, do not add these fluxes) and include the effects from the higher resonances.

From \figu{xsec}, one can read off that there are two classes of resonances: The first peak in \figu{xsec} is dominated by the $\Delta(1232)$-resonance (lower resonance -- LR), whereas the higher resonances (HR) contribute at larger energies. In addition, the kinematics (\cf, $\chi$  in \Tab~\ref{tab:simmodela}) and the multiplicities (\cf, \Tab~\ref{tab:numbers}) are very different. For example, protons and neutrons interactions via the $\Delta(1232)$-resonance happen through  only one interaction type, and produce either only $\pi^+$ or only $\pi^-$ (and $\pi^0$). For the higher resonances, the pions are produced in all pion charges.

\begin{figure}[t!]
\begin{center}
\includegraphics[width=10cm]{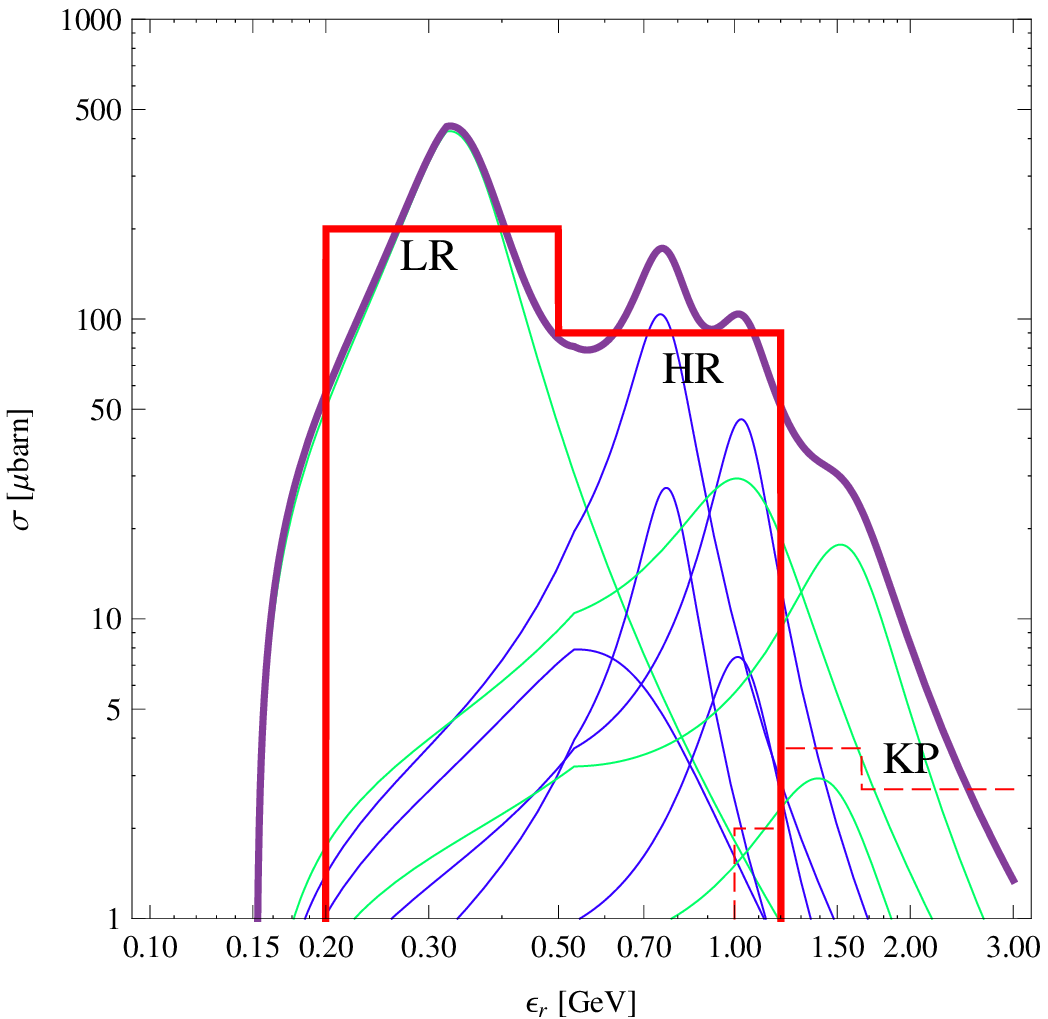}
\end{center}
\mycaption{\label{fig:res} Cross section for the resonances as a function of $\epsilon_r$ (thick curve). Green (light gray) resonances are $\Delta$-resonances, blue (dark gray resonances) are $N$-resonances. The total resonance cross section is shown as thick curve. Our simplified model~B is depicted by the red (gray) boxes, where the interaction types are labeled. The dashed curve refers to our simplified cross section for kaon production ``KP''.}
\end{figure}

\begin{table}
\begin{center}
\begin{tabular}{clr|ccc|c|ccc|c|r}
\hline
 IT & $\epsilon_r$-range [GeV] & $\sigma$ [$\mu$barn]  & \multicolumn{4}{c|}{Initial proton} & \multicolumn{4}{c|}{Initial neutron} & \\
\hline
\hline
 & &  & $M_{\pi^0}$ & $M_{\pi^+}$ & $M_{\pi^-}$ & $M_n$ & $M_{\pi^0}$ & $M_{\pi^+}$ & $M_{\pi^-}$  & $M_n$ &  K \\
 & & & $\chi_{\pi^0}$ & $\chi_{\pi^+}$ & $\chi_{\pi^-}$ & $M_p$ & $\chi_{\pi^0}$ & $\chi_{\pi^+}$ & $\chi_{\pi^-}$ & $M_p$  &  \\
\hline
LR & $0.2 \hdots  0.5 $ & 200 & 2/3 & 1/3 & - & 1/3 &  2/3 & - & 1/3  & 2/3 & 0.22 \\
 & &  & 0.22 & 0.22 & - & 2/3 & 0.22 &  -  & 0.22  & 1/3 \\
\hline
HR & $0.5 \hdots  1.2 $ & 90 & 0.47 & 0.77 & 0.34 & 0.43 & 0.47 & 0.34 & 0.77 & 0.57 & 0.39 \\
 & &  & 0.26 & 0.25 & 0.22 & 0.57 & 0.26 & 0.22 & 0.25 & 0.43   \\
\hline
\end{tabular}
\end{center}
\mycaption{\label{tab:simmodelb} Parameters for the resonant pion production in our simplified treatment~B.
}
\end{table}

For the resonances, we define two interaction types LR and HR, as shown in \figu{res} (boxes). The interaction type LR corresponds to $\Delta_1(1232)$, whereas the interaction type HR contains the higher resonances.
The properties of these interaction types are summarized in \Tab~\ref{tab:simmodelb}.

The surface area covered by these cross sections, corresponding to the $\hat \Gamma$ in simplified model~A, is chosen for LR and HR such that the pion spectra match to these of the previous section for typical power law spectra (with spectral indices of about two). Of course, there can never be an exact match, because the contributions from the individual resonances depend on the spectral index. However, as we will demonstrate, this estimate is good enough for our purposes.

The averaged numbers for the multiplicities and inelasticities for the higher resonances interaction type  are estimated from the interaction rate \equ{pgamma2} by assuming that all resonances contribute simultaneously and weighting by the interaction type-dependent part $\epsilon_{r}^{\mathrm{IT}, 0} \,  \hat\Gamma^{\mathrm{IT}}  \, B^{\mathrm{IT}}_{\mathrm{out}}$ (\cf, \equ{irates} using \equ{fa}; \cf, \App~\ref{app:others}). By the same procedure we obtain the average $\chi$-values from \equ{response}. Note that the $\pi^-$ (for proton interactions) are, in average, reconstructed at lower energies, because these are mostly produced by interaction type~2, which has smaller $\chi$-values (\cf, \Tab~\ref{tab:simmodela}). In addition, note that the total number of charged pions is, in fact, close to one per interaction here, and the total number of pions close to 1.5, since in interaction types~2 and~3 more than one pion is produced.

The pion spectra are then computed from \equ{prodsimple} with
\begin{eqnarray}
f^{\mathrm{LR}} & = & \left\{
\begin{array}{ll}
0 & 2y < 0.2 \, \mathrm{GeV} \\
200 \, \mu \mathrm{barn} \, \left(1- \frac{(0.2 \, \mathrm{GeV})^2}{(2y)^2} \right) & 0.2 \, \mathrm{GeV} \le 2y < 0.5 \, \mathrm{GeV}\\
200 \, \mu \mathrm{barn} \, \left(\frac{(0.5 \, \mathrm{GeV})^2-(0.2 \,  \mathrm{GeV})^2}{(2y)^2} \right) & 2y \ge 0.5 \, \mathrm{GeV}\\
\end{array} \right. \\
f^{\mathrm{HR}} & = & \left\{
\begin{array}{ll}
0 & 2y < 0.5 \, \mathrm{GeV} \\
90 \, \mu \mathrm{barn} \, \left(1- \frac{(0.5 \, \mathrm{GeV})^2}{(2y)^2} \right) & 0.5 \, \mathrm{GeV} \le 2y < 1.2 \, \mathrm{GeV}\\
90 \, \mu \mathrm{barn} \,  \left(\frac{(1.2 \, \mathrm{GeV})^2-(0.5 \,  \mathrm{GeV})^2}{(2y)^2} \right) & 2y \ge 1.2 \, \mathrm{GeV} \, ,
\end{array}
  \right.
\end{eqnarray}
as it can be shown from \equ{split}.
Here $M_\pi^{\mathrm{IT}}$  and $\chi^{\mathrm{IT}}$ (needed in \equ{prodsimple}) are chosen according to interaction type, initial proton or neutron, and final pion, as given in \Tab~\ref{tab:simmodelb}.

\subsubsection{Comparison of resonance response functions}

\begin{figure}[t!]
\begin{center}
\includegraphics[width=\textwidth]{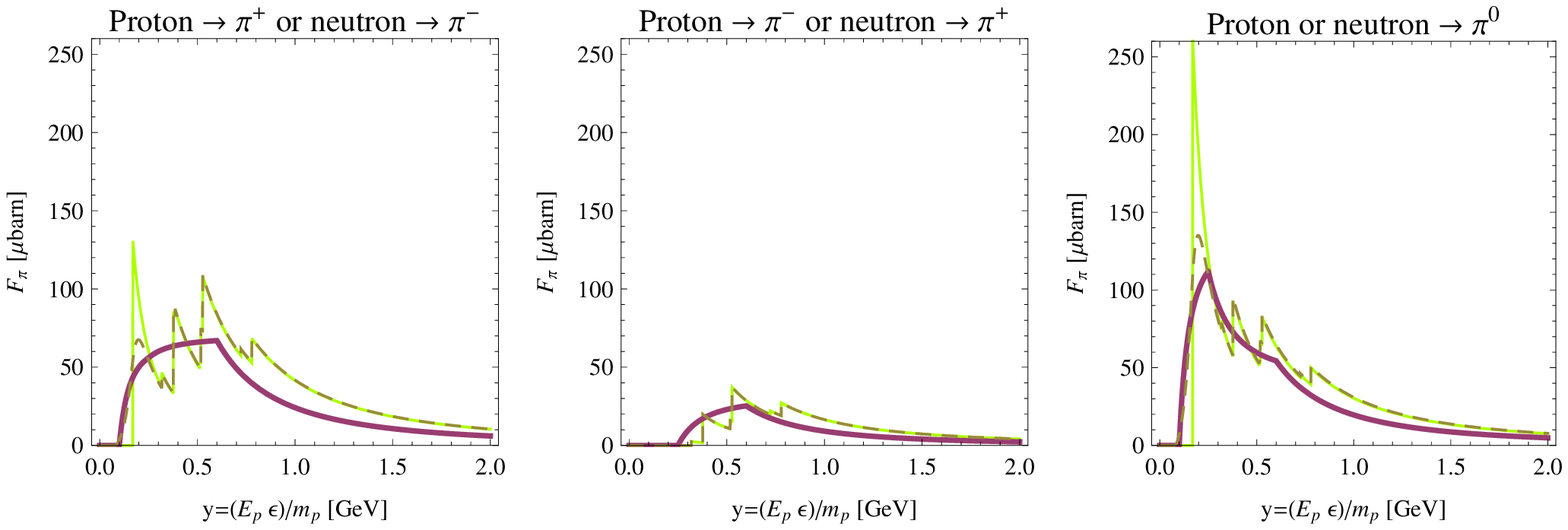}
\end{center}
\mycaption{\label{fig:fres} Response function $F_\pi(y)\equiv \sum_{\mathrm{IT}}  M_\pi^{\mathrm{IT}} \, f^{\mathrm{IT}}(y)$ summed over the resonances only. Here simplified model Sim-A (thin solid curves) is compared with simplified model Sim-B (thick curves). The thin dashed curves correspond to Sim-A with the full Breit-Wigner form of the interaction type $\Delta_1(1232)$ only.  }
\end{figure}

In \figu{fres}, we compare the response function $F_\pi(y) \equiv \sum_{\mathrm{IT}}  M_\pi^{\mathrm{IT}} \, f^{\mathrm{IT}}(y)$ between simplified model Sim-A (thin solid curves) and Sim-B (thick curves), summed over the resonances. This function is proportional to the number of produced pions of a certain species as a function of $y$, whereas the $x$-dependent part in \equ{split} describes the energy at which the pions are found. Obviously, the function is much smoother for Sim-B than for Sim-A. Because of only a few contributing interaction types and the smoothness of the function, the evaluation will be much faster. Once the photon spectrum is folded in, the contributions of both response functions will be very similar.

Note that model Sim-B includes the part of the $\Delta(1232)$-resonance below the peak. This can be seen by comparing with the dashed curve, which represents model Sim-A but the interaction type $\Delta_1(1232)$ replaced by the full Breit-Wigner-form.

\subsection{Direct production}
\label{sec:directsimple}

\begin{table}[t!]
\begin{center}
\begin{tabular}{lrrrr|ccc|ccc}
\hline
& & & & & \multicolumn{3}{c|}{Initial proton} & \multicolumn{3}{c}{Initial neutron} \\
IT & $\epsilon_{\mathrm{min}}^{\mathrm{IT}}$ [GeV] & $\epsilon_{\mathrm{max}}^{\mathrm{IT}}$ [GeV] & $\chi$   & K & $M_{\pi^0}$ & $M_{\pi^+}$ & $M_{\pi^-}$ &  $M_{\pi^0}$ & $M_{\pi^+}$ & $M_{\pi^-}$\\
\hline
T$_{\text{1L}}$ & $0.17$ &  $0.56 $ & 0.13 & 0.13 & - & 1 & - & - & - & 1 \\
T$_{\text{1M}}$ & $0.56$ &  $10 $ & 0.05 &  0.05 & - & 1 & - & - & - & 1 \\
T$_{\text{1H}}$ & $10$ &  $\infty$  & 0.001 &  0.001 & - & 1 & - & - & - & 1 \\
\hline
T$_{\text{2aL}}$ & $0.4$ &  $1.58 $ & 0.08 & 0.28  &   - & $\frac{1}{4}$ & $\frac{3}{4}$ &  - & $\frac{3}{4}$ & $\frac{1}{4}$  \\
T$_{\text{2aM}}$ & $1.58$ &  $10 $ & 0.02 & 0.22 &   - & $\frac{1}{4}$ & $\frac{3}{4}$ &  - & $\frac{3}{4}$ & $\frac{1}{4}$  \\
T$_{\text{2aH}}$ & $10$ & $\infty$ & 0.001 &  0.201 &   - & $\frac{1}{4}$ & $\frac{3}{4}$ &  - & $\frac{3}{4}$ & $\frac{1}{4}$  \\
\hline
T$_{\text{2b}}$ & $ 0.4$ & $\infty$ & 0.2 &  -  &  $\frac{1}{6}$ & $\frac{3}{4}$ & $\frac{1}{12}$ & $\frac{1}{6}$ & $\frac{1}{12}$ & $\frac{3}{4}$  \\
\hline
\end{tabular}
\end{center}
\mycaption{\label{tab:simdirectmodel} Parameters for the direct pion production in our simplified treatment. The inelasticity for T$_{\text{2b}}$ is included in T$_{\text{2a}}$.}
\end{table}

For direct production, we also follow the approach in \Sec~\ref{sec:rsimple}. However, this interaction type is tricky, since the kinematics function $\chi^\text{IT}(\epsilon_r)$ is strongly dependent on the interaction energy (see \App~\ref{app:directkin} for details). This implies that the $\delta$-distribution in \equ{nabsimple} is not a good approximation and needs to be replaced by a broader distribution function. In addition, the distribution of scattering angles will lead to smearing effects. In this case, it turns out to be a good approximation to split the direct production in different interaction types with different characteristic values of $\chi^\text{IT}$ as a function of $\epsilon_r$, which simulates such a broad distribution after the integration of the input spectra. We define three interaction types $T_{\text{1L}}$, $T_{\text{1M}}$, and $T_{\text{1H}}$ for direct one pion production and four for two pion production, namely  $T_{\text{2aL}}$, $T_{\text{2aM}}$, and $T_{\text{2aH}}$ for the first pion, and $T_{\text{2b}}$ for the second pion. The interaction types are shown in \Tab~\ref{tab:simdirectmodel}, and the names correspond to \Tab~\ref{tab:numbers}, split into low (L), medium (M), and high (H) energy parts in $\epsilon_r$, respectively, limited by $\epsilon_{\mathrm{min}}^{\mathrm{IT}}$ and $\epsilon_{\mathrm{max}}^{\mathrm{IT}}$. After this splitting, the additional effect of the scattering angle smearing turns out to be small, as we will demonstrate later.
We obtain the function $f^{\mathrm{IT}}$ for these interaction types from \equ{split} as
\begin{equation}
 f^{\mathrm{IT}}  =  \left\{
\begin{array}{ll}
0 & 2y < \epsilon_{\mathrm{min}}^{\mathrm{IT}} \\
\frac{1}{2y^2}\left(I^{\mathrm{IT}}(2y)-I^{\mathrm{IT}}(\epsilon_{\mathrm{min}}^{\mathrm{IT}})\right) & \epsilon_{\mathrm{min}}^{\mathrm{IT}} \le 2y < \epsilon_{\mathrm{max}}^{\mathrm{IT}}\\
\frac{1}{2y^2}\left(I^{\mathrm{IT}}(\epsilon_{\mathrm{max}}^{\mathrm{IT}})-I^{\mathrm{IT}}(\epsilon_{\mathrm{min}}^{\mathrm{IT}})\right) & 2y \ge \epsilon_{\mathrm{max}}^{\mathrm{IT}}\\
\end{array}
\right.\\
\end{equation}
with $I^{\mathrm{IT}}(2y)$  re-parameterized using $x \equiv \log_{10}(y/\mathrm{GeV})$:
\begin{eqnarray}
I^{\mathrm{T}_{1}}(2y) & = & \left\{
\begin{array}{ll}
0 \\
\qquad 2 y < 0.17 \, \mathrm{GeV} \\
35.9533+84.0859 x+110.765 x^2+102.728 x^3+40.4699 x^4 \label{equ:d1}  \\ \qquad 0.17\, \mathrm{GeV}< 2 y<0.96  \, \mathrm{GeV} \\
30.2004 + 40.5478 x + 2.03074 x^2 -
 0.387884 x^3 + 0.025044 x^4  \\
\qquad 2y > 0.96 \, \mathrm{GeV}
\end{array}
\right.
\\
I^{\mathrm{T}_{2}}(2y) & = &
 \left\{
\begin{array}{ll}
0 & 2 y < 0.4 \, \mathrm{GeV} \\
-3.4083+\frac{16.2864}{2\,y}+40.7160 \ln{(2 y)}
& 2 y \ge 0.4 \, \mathrm{GeV} \\
\end{array}
\right.
\label{equ:d2}
\end{eqnarray}
Note that $\epsilon_{\mathrm{min}}^{\mathrm{IT}}$, $\epsilon_{\mathrm{max}}^{\mathrm{IT}}$, the multiplicities and $\chi^{\mathrm{IT}}$ can be read off from  \Tab~\ref{tab:simdirectmodel} for the different interaction types.
The integral values $I^{\mathrm{IT}}$ are the same for interaction types $\mathrm{T_{1L}}$, $\mathrm{T_{1M}}$, $\mathrm{T_{1H}}$ (\equ{d1}), and for interaction types $\mathrm{T_{2aL}}$, $\mathrm{T_{2aM}}$, $\mathrm{T_{2aH}}$, and $\mathrm{T_{2b}}$ (\equ{d2}). In \equ{prodsimple}, all the interaction types in \Tab~\ref{tab:simdirectmodel} have to be summed over.

\subsection{Multi-pion production}

Here we show two different approaches to the multi-pion channel. The first, simplified approach will later be shown as model Sim-C, the second, more refined approach will be used in all other models.

\subsubsection{Simplified kinematics}
\label{sec:mp1}

We start with the simplest example for multi-pion production, for which we follow \Sec~\ref{sec:kinsec}. We assume $\chi^{\text{Multi}-\pi} \simeq 0.2$, \ie,  the response function factorizes as $R^{\text{Multi}-\pi} = \delta(x-0.2) \, M_\pi^\mathrm{Multi-\pi}  \, f^{\mathrm{Multi-\pi}} ( y )$.
We obtain from \equ{split} and \equ{prodsimple}
\begin{equation}
Q_{\pi}^{\mathrm{Multi-\pi}}(E_\pi) = N_p(5 E_\pi) \, \frac{m_p}{E_\pi} \, \int\limits_{\epsilon_{\mathrm{th}}/2}^\infty
d y \,  M_\pi^\mathrm{Multi-\pi}  \, f^{\mathrm{Multi-\pi}} ( y ) \, n_\gamma \left( \frac{m_p y}{5 E_\pi} \right) \,  .
\label{equ:multips}
\end{equation}
with
\begin{equation}
f^{\mathrm{Multi-\pi}}(y) \equiv \frac{1}{2 y^2} \int\limits_{\epsilon_{\mathrm{th}}}^{2y} d \epsilon_r \, \epsilon_r\,  \sigma^{\mathrm{Multi-\pi}}(\epsilon_r)  \, .
\label{equ:multipg}
\end{equation}
As described in \Sec~\ref{sec:photo}, we assume that $M_{\pi^0}=1$ ($M_{\pi^0}=1$), $M_{\pi^+}=1.2$ ($M_{\pi^+}=0.85$), and $M_{\pi^-} = 0.85$ ($M_{\pi^-} =1.2$) for initial protons (neutrons).
The function $f^{\mathrm{Multi-\pi}}(y)$ can be obtained using the sum of \equ{mp1} and \equ{mp2}. We numerically integrate it and re-parameterize it with $x=\log_{10}(y/\mathrm{GeV})$ by
{\small
\begin{equation}
\frac{\hat f(x)_{\pi^{+,0}}}{\mu \text{barn}} = \left\{
\begin{array}{ll}
0 &   2 y<0.5  \, \mathrm{GeV} \\
87.5538 + 120.894 x - 98.4187 x^2 -  59.6965 x^3 + 67.2251 x^4  &  0.5 \, \mathrm{GeV} \le 2 y \le 14 \, \mathrm{GeV} \\
131.839 - 25.3296 x + 10.612 x^2 - 0.858307 x^3 + 0.0493614 x^4  &   2y > 14 \, \mathrm{GeV}
\end{array}
\right.
\label{equ:mp}
\end{equation}
} 
for  $\pi^0$,  and $\pi^+$ (initial proton) or  $\pi^-$ (initial neutron), and
{\small
\begin{equation}
\frac{\hat f(x)_{\pi^{-}}}{\mu \text{barn}} = \left\{
\begin{array}{ll}
0 &   2 y< 1  \, \mathrm{GeV} \\
73.9037 + 187.526 x - 161.587 x^2 - 206.268 x^3 + \\
\qquad + 354.02 x^4 -
 129.759 x^5
  &  1 \, \mathrm{GeV} \le 2 y \le 10 \, \mathrm{GeV} \\
131.839 - 25.3296 x + 10.612 x^2 - 0.858307 x^3 + 0.0493614 x^4  &   2y > 10 \, \mathrm{GeV}
\end{array}
\right.
\label{equ:mpminus}
\end{equation}
} 
for $\pi^-$ (initial proton) or  $\pi^+$ (initial neutron). Note that the different function for $\pi^-$ comes from the different threshold below which we have set the cross section to zero, as discussed in \Sec~\ref{sec:kinsec}.
For $y \gg 10^4 \, \mathrm{GeV}$, this function still increases as extrapolation of  \figu{xsec}  (the cross section is not measured in that energy range).

\subsubsection{Kinematics simulated from SOPHIA}
\label{sec:mp2}

\begin{table}[t!]
\begin{center}
\begin{tabular}{lrrrrr|ccc|ccc}
\hline
& & & &  & & \multicolumn{3}{c|}{Initial proton} & \multicolumn{3}{c}{Initial neutron} \\
IT & $\epsilon_{\mathrm{min}}^{\mathrm{IT}}$ [GeV] & $\epsilon_{\mathrm{max}}^{\mathrm{IT}}$ [GeV] &  $\sigma$ &$\chi$  &  K & $M_{\pi^0}$ & $M_{\pi^+}$ & $M_{\pi^-}$ &  $M_{\pi^0}$ & $M_{\pi^+}$ & $M_{\pi^-}$\\
\hline
M$_{\text{1L}}$ & 0.5 & 0.9 & 60 & 0.1 & 0.27  & 0.32 & 0.34 & 0.04 & 0.32 & 0.04 & 0.34 \\
M$_\text{1H}$ & 0.5 & 0.9 & 60 &  0.4 & -- & 0.17 & 0.29 & 0.05 & 0.17 & 0.05 & 0.29\\
M$_\text{2L}$ & 0.9& 1.5 & 85 & 0.15 & 0.34 & 0.42& 0.31& 0.07 & 0.42& 0.07& 0.31\\
M$_\text{2H}$ & 0.9& 1.5& 85 & 0.35&-- & 0.19& 0.35& 0.08& 0.19& 0.08& 0.35\\
M$_\text{3L}$& 1.5& 5.0& 120 & 0.15& 0.39 & 0.59& 0.57& 0.30&   0.59& 0.30& 0.57\\
M$_\text{3H}$& 1.5&  5.0& 120 & 0.35&--& 0.16& 0.21& 0.13& 0.16& 0.13& 0.21 \\
M$_\text{4L}$& 5.0&   50& 120& 0.07& 0.49 & 1.38& 1.37& 1.11& 1.38& 1.11& 1.37\\
M$_\text{4H}$& 5.0& 50& 120 & 0.35& --& 0.16& 0.25& 0.23&0.16& 0.23& 0.25\\
M$_\text{5L}$& 50& 500& 120 & 0.02& 0.45 & 3.01& 2.86& 2.64&3.01& 2.64& 2.86\\
M$_\text{5H}$& 50& 500& 120 & 0.5&-- & 0.20& 0.21& 0.14&0.20& 0.14& 0.21\\
M$_\text{6L}$& 500& 5000& 120 & 0.007& 0.44 & 5.13& 4.68& 4.57&5.13& 4.57& 4.68\\
M$_\text{6H}$& 500& 5000& 120 & 0.5& --& 0.27& 0.29& 0.12&0.27& 0.12& 0.29\\
M$_\text{7L}$& 5000& $\infty$ & 120 & 0.002& 0.44 & 7.59& 6.80& 6.65&7.59& 6.65& 6.80\\
M$_\text{7H}$& 5000& $\infty$ & 120 & 0.6& --& 0.26& 0.27& 0.13 & 0.26& 0.13& 0.27\\
\hline
\end{tabular}
\end{center}
\mycaption{\label{tab:mtype} Parameters for the multi-pion production in our SOPHIA-based treatment. The inelasticities for the H-sample are included in the L-sample.}
\end{table}

Similar to the direct production, the $\delta$-distribution in \equ{nabsimple} is not a good approximation and needs to be replaced by a broader distribution function. We again solve this by splitting the multi-pion production in different interaction types with different characteristic values of $\chi^\text{IT}$, which simulates such a broad distribution after the integration of the input spectra. Compared to the direct production, the cross section can be parameterized relatively easily by step functions. However, the pion multiplicities in fact increase with energy. This means that the higher the interaction energy, the more pions are produced, which (in average) are found at lower energies. In addition, the final energy distribution functions are broad.

We solve this strongly scale-dependent behavior by dividing the  $\epsilon_r$-range in seven interaction types M$_i$, each with a particular average cross section and average pion multiplicities, which we directly take from SOPHIA. In addition, we split the pions in each of these samples in a lower energy (L) and higher energy (H) part, which are reconstructed at different values of $\chi^{\mathrm{IT}}$ to simulate the broadth of the distributions within the same $\epsilon_r$. Typically, the $L$-sample corresponds to the peak of the distribution (at least for high energies), whereas the H-sample simulates the tail of the distribution. The splitting of the multiplicities can be performed automatically once a splitting point is defined. We choose the $\chi$-values of the interaction types to simulate the peaks of the distribution and to reproduce the total energy going into pions in SOPHIA. This treatment is, of course, not extremely accurate, but it allows to use the fast approach in \equ{split} while obtaining accuracy at high energies.

The function $f^\mathrm{M}$ can be easily calculated from \equ{split} since the cross section is assumed to be constant within each IT:
\begin{equation}
 f^{\mathrm{M}_i}  =  \left\{
\begin{array}{ll}
0 & 2y < \epsilon_{\mathrm{min}}^{\mathrm{M}_i} \\
\frac{\sigma^{\mathrm{M}_i}}{(2y)^2}\left((2y)^2-(\epsilon_{\mathrm{min}}^{\mathrm{M}_i})^2 \right) & \epsilon_{\mathrm{min}}^{\mathrm{M}_i} \le 2y < \epsilon_{\mathrm{max}}^{\mathrm{M}_i}\\
\frac{\sigma^{\mathrm{M}_i}}{(2y)^2}\left((\epsilon_{\mathrm{max}}^{\mathrm{M}_i})^2-(\epsilon_{\mathrm{min}}^{\mathrm{M}_i})^2 \right) & 2y \ge \epsilon_{\mathrm{max}}^{\mathrm{M}_i}\\
\end{array}
\right.\\
\end{equation}
The interaction types are listed in \Tab~\ref{tab:mtype}, where the parameters for this equation can be found, as well as the $\chi$ and $K$ values and multiplicities for the individual interaction types.

\subsection{Kaon production}

Here we include $K^+$ production into our simplified model. If needed, other (sub-leading) channels can be added to our framework as well (such as $K^-$ production if the neutrino-antineutrino ratio in the higher energy regime is studied). This example should also serve as illustration how to add processes to our model. Note that this implementation, however, is more primitive than the pion production above.

The production of $K^+$ in photohadronic interactions cannot be assigned to a single resonance, but comes from a number of resonances mainly decaying into $K^+$ and $\Lambda$ or $\Sigma^0$ (with relatively similar masses, which means that we can neglect the mass difference). In addition, at high interaction energies, multi-fragmentation, similar to multi-pion production, contributes significantly.
The production cross section up to $\epsilon_r \simeq 2 \, \mathrm{GeV}$ has been measured~\citep{Tran:1998qw}. For higher energies, where no data are available, one may use extrapolations from models~\citep{Lee:1999kd,Asano:2006zzb}. Approximating the data in \citet{Lee:1999kd} and extrapolating according to \citet{Lee:1999kd}, we define an interaction type ``KP'' and model the total $K^+$ production cross section as (\cf, dashed curve in \figu{res})
\begin{eqnarray}
\sigma^{\mathrm{KP}} & = & \left\{
\begin{array}{ll}
0 & \epsilon_r < 1.0 \, \mathrm{GeV} \\
2.0 \, \mu \mathrm{barn}  & 1.0 \, \mathrm{GeV}\le \epsilon_r < 1.2 \, \mathrm{GeV}\\
3.7 \, \mu \mathrm{barn}   & 1.2 \, \mathrm{GeV}\le  \epsilon_r < 1.65 \, \mathrm{GeV}\\
2.7 \, \mu \mathrm{barn}    & \epsilon_r \ge 1.65 \, \mathrm{GeV} \\
\end{array}
  \right.
\end{eqnarray}
Accordingly, we obtain the response function from \equ{split} with $M_{K^+}=1$:
\begin{eqnarray}
f^{\mathrm{KP}} & = & \left\{
\begin{array}{ll}
0 & 2y < 1.0 \, \mathrm{GeV} \\
2.0 \, \mu \mathrm{barn}  \, \left(1- \frac{1.0 \, \mathrm{GeV}^2}{(2y)^2} \right) & 1.0 \, \mathrm{GeV} \le 2y < 1.2 \, \mathrm{GeV}\\
3.7 \, \mu \mathrm{barn}  \, \left(1- \frac{0.3 \, \mathrm{GeV}^2}{(2y)^2} \right) & 1.2 \, \mathrm{GeV} \le 2y < 1.65 \, \mathrm{GeV}\\
2.7 \, \mu \mathrm{barn}  \, \left(1- \frac{0.16 \, \mathrm{GeV}^2}{(2y)^2} \right)  & 2y \ge 1.65 \, \mathrm{GeV}\\
\end{array}
  \right.
\end{eqnarray}
From \Sec~\ref{sec:kinsec}, we have $\chi_{K^+} \simeq 0.35$ (computed for $\epsilon_r \simeq 1.4 \, \mathrm{GeV} $ at the peak). The effects on primary cooling and secondary re-injection in \App~\ref{app:others} are negligible. Note that the absolute normalization of the kaons at high energies crucially depends on the extrapolation of the cross sections to high energies. Our kaon production is about a factor of two smaller at higher energies than in \citet{Lipari:2007su}, since we assume the cross section to be constant at high energies.

\subsection{Comparison of individual contributions}

\begin{figure}[t]
\begin{center}
\includegraphics[width=\textwidth]{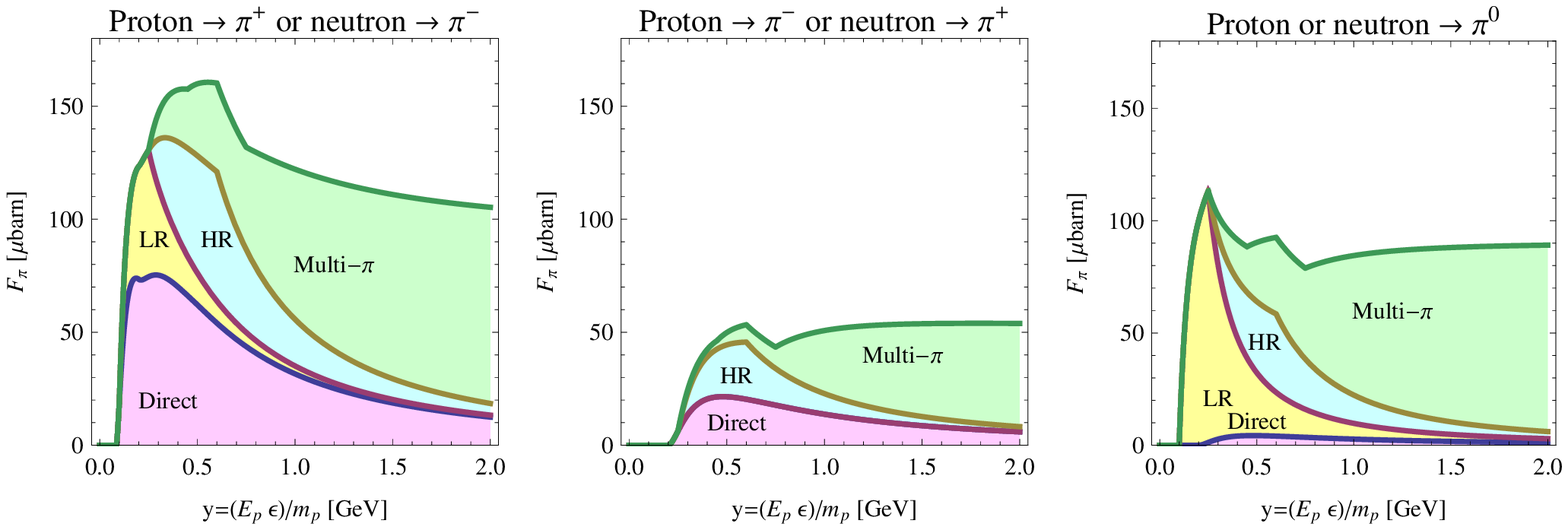}
\mycaption{\label{fig:resmodelb} The function $F_\pi(y) \equiv \sum_{\mathrm{IT}}  M_\pi^{\mathrm{IT}} \, f^{\mathrm{IT}}(y)$ for pion production (resonance treatment from simplified model~B), with the individual contributions marked.}
\end{center}
\end{figure}
In \figu{resmodelb}, we show the individual contributions to $F_\pi(y) \equiv \sum_{\mathrm{IT}}  M_\pi^{\mathrm{IT}} \, f^{\mathrm{IT}}(y)$ for pion production as a function of $y$, which is proportional to the maximal available center-of-mass energy.  This quantity describes the total number of pions of a particular species produced as a function of $y$, including multiplicity and cross section. However, it does not include the energy where the pions are found.
For initial protons, $\pi^+$ are most abundantly produced because of the direct production contribution, followed by $\pi^0$ and then $\pi^-$. For $\pi^+$ production close to the threshold, the direct production and resonances are most important, whereas for  $\pi^0$ production, the direct production hardly has any impact, and the resonances dominate. For $\pi^-$ production, at the threshold all processes contribute almost equally, only the lower resonance does not take part. For larger $y$, in all cases  multi-pion production dominates.

As far as the relative contribution is concerned, thanks to the direct production, the $\pi^+$ are always produced most abundantly. As one can read off from \figu{resmodelb}, this statement is independent of the photon or proton spectra used, because the response function for $\pi^+$ is always larger than the one for $\pi^0$, in a large part of the energy range even about 50\% larger. This means that for arbitrary proton and photon spectra, thanks to the direct production, the $\pi^+$:$\pi^0$ ratio lies between about 1:1 and 3:2, whereas the $\Delta(1232)$-approximation in \equ{ds} predicts 1:2. In fact, one can show that the minimum of the charged to neutral pion ratio with respect to $y$ is
\begin{equation}
\underset{y}{\mathrm{min}} \frac{F_{\pi^+}(y)+F_{\pi^-}(y)}{F_{\pi^0}(y)} \simeq 1.2  \,
\label{equ:Fmin}
\end{equation}
{\em for arbitrary input spectra}; see~\cite{2000NuPhS..80C0810M} for a specific photon field.\footnote{As mentioned above, $F_\pi(y)$ does not include the reconstructed energy of the pions. Since, however, the pions of the different species are found in similar energy ranges, this statement also roughly applies to the energy deposited into the different species.} From this equation, any neutrino flux computed with the $\Delta(1232)$-approximation and normalized to the observed photon spectrum, if mainly coming from $\pi^0$ decays, is underestimated by a factor of at least $1.2/0.5 = 2.4$, \ie, the neutrino flux should be about a factor of 2.4 larger. Even the often used approximation that 50\% of all photohadronic interactions result in charged pions underestimates the neutrino flux by at least 20\%. These numbers are to be interpreted as lower limits for the neutrino flux underestimation, the exact values depend on the input spectra and may be even higher.

Of course, the overall impact of the individual contributions depends on the proton and photon spectra the response function is to be folded with. In order to check the impact of the individual contributions on typical AGN, GRB or high temperature black body (BB) spectra,
we define three benchmark spectra in \App~\ref{app:benchmarks}. Note that all of the spectra shown in this work are given in the SRF.
\begin{figure}[]
\begin{center}
\includegraphics[width=\textwidth]{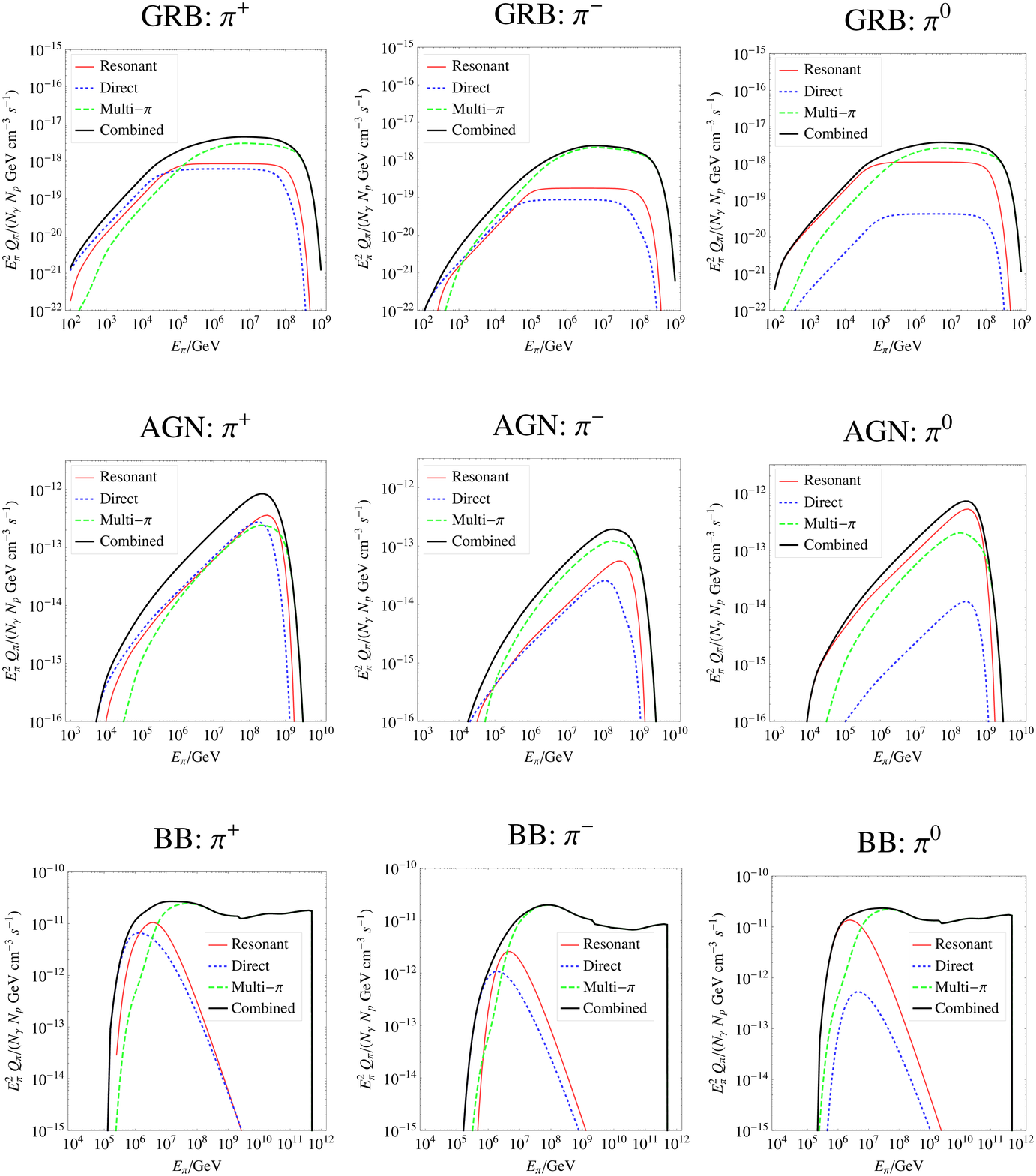}
\end{center}
\mycaption{\label{fig:piind} Contributions of resonant (thin solid), direct (dotted) and multi-pion (dashed) production for $\pi^+$, $\pi^-$ and $\pi^0$ spectra (from left to right) for proton-photon interactions. The upper panels are for the GRB benchmark, the middle ones for the AGN benchmark  and the lower ones for the BB benchmark (see \App~\ref{app:benchmarks}). Computed with model Sim-B.}
\end{figure}
In \figu{piind}, we show the  $\pi^+$, $\pi^-$, and $\pi^0$ spectra for these benchmarks. The upper panels are for the GRB benchmark, the middle ones for the AGN benchmark and the lower ones for the BB benchmark. For the GRB benchmark, direct production dominates at low energies for the $\pi^+$ production, whereas the $\pi^0$ production at low energies is determined by the resonances. The $\pi^-$ production is dominated at all energies by the multi-pion production, such as the other spectra at high energies. Therefore, all different processes are important, but they contribute entirely different to the different pion polarities.
One can also read off from this figure, that the characteristic shape of the GRB pion or neutrino spectra often shown in the literature (resonance curves), which is flat in the middle energy range, is tilted upwards due to multi-pion production.
For the AGN benchmark (middle panels), the $\pi^0$ production is governed by the resonances in the whole energy range, the $\pi^-$ production by the multi-pion production, and the $\pi^+$ production by similar contributions of all processes including the direct production.  In this case, the $\Delta$-resonance approximation is more accurate in terms of the shapes of the spectra. However, other processes quantitatively contribute as well.
For the BB benchmark (lower panels), the $\pi$ production is dominated by the multi-pion production for most of the energies. Only in the low energy region, the charged pion production is governed by the direct and the resonant processes, and the $\pi^0$ production by the resonances. In this case, ignoring the high energy processes would lead to clearly misleading results.

\begin{figure}[t!]
\begin{center}
\includegraphics[width=\textwidth]{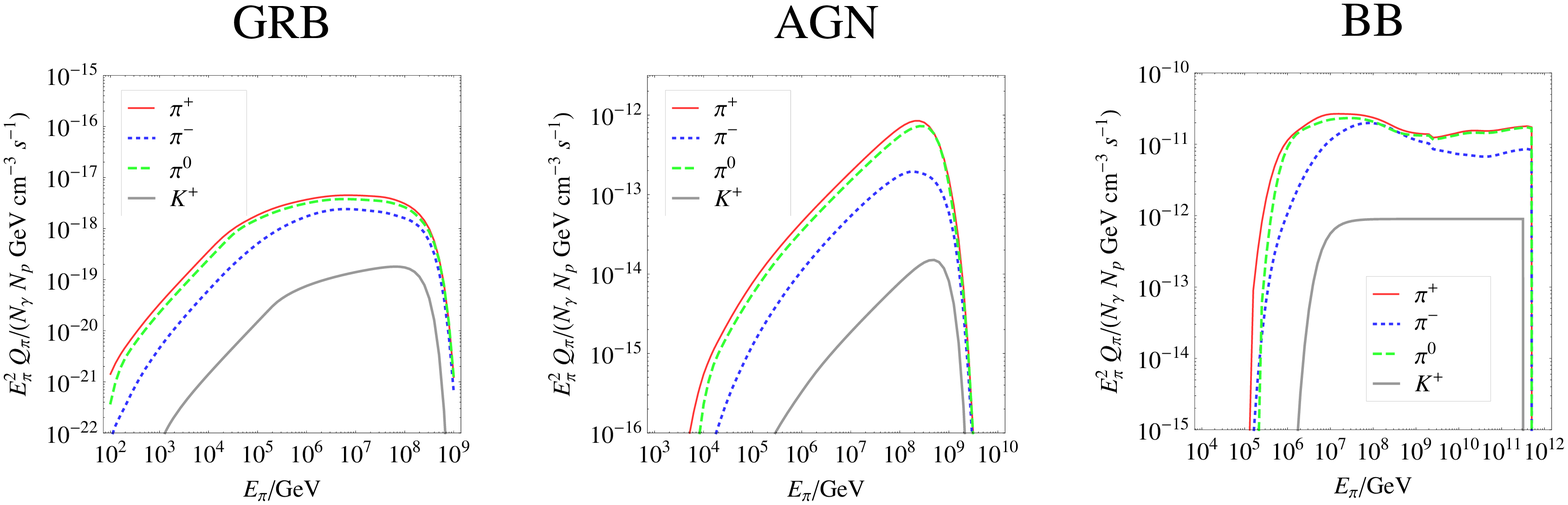}
\end{center}
\mycaption{\label{fig:pispectra} Comparison among the $\pi^+$ (upper curve), $\pi^0$ (middle curve), and $\pi^-$ (lower curve) spectra for GRB (left), AGN (middle) and BB (right) benchmark. The grey curve shows in addition the $K^+$ spectrum. Computed with model Sim-B.}
\end{figure}

As far as the  comparison among the different polarities is concerned, see \figu{pispectra}. For all spectra, the $\pi^+$ are always most abundantly produced, as predicted above, followed by $\pi^0$ and then $\pi^-$. For the GRB benchmark at high energies, where the multi-pion production dominates, the spectra are closest to each other. At low energies, there are significantly less $\pi^-$ produced than the other two polarities. However, note that, thanks to the multi-pion production, the $\pi^-$ are only suppressed by a factor of a few. The kaons, on the other hand, contribute about one to two orders less to the total meson fluxes. Nevertheless, there can be interesting effects in the high energy regime if cooling effects are present. For the AGN benchmark, because of the lower maximal proton energy times photon energy, even for large energies the $\pi^-$ are strongly suppressed. Otherwise, the result is qualitatively similar. For the BB spectrum we can nicely see the effect of lower multiplicities of $\pi^-$ compared to $\pi^+$ and $\pi^0$ for high interaction energies (see \Tab~\ref{tab:mtype}) in the spectrum at energies higher than $10^9\,\mathrm{GeV}$. Otherwise, we have similar results as for the GRB benchmark.

\section{Comparison with SOPHIA}
\label{sec:comparison}

Here we compare the results of our simplified models with each other and with SOPHIA. First, we focus on the primaries produced in the photohadronic interactions, mostly the pions. Then we compute and compare the neutrino spectra. In all cases, we use initial protons for the photohadronic interactions.

\subsection{Pion spectra}

\begin{table}[t]
\begin{center}
\begin{small}
\begin{tabular}{lp{6.5cm}p{4.5cm}p{2.3cm}}
\hline
Abbrev. & Description & Complexity & \Sec/Refs. \\
\hline
SOPHIA & SOPHIA software with full kinematics & Monte Carlo method, $\sim~3000\times  T_\mathrm{Sim-C}$ & \citet{Mucke:1999yb} \\
BW & Resonances with full Breit-Wigner description; direct production with full response function; simple kinematics & Double integration, 45 IT, $\sim~120\times  T_\mathrm{Sim-C}$ & \Sec~\ref{sec:photo} incl. multi-pion from \Sec~\ref{sec:mp2} \\
Sim-A &  Simplified model with  factorized response function and resonance treatment~A ($\delta$-function approximation)  & Single integration, 49 IT, $\sim~(3-4)\times T_\mathrm{Sim-C}$ & \Sec~\ref{sec:simplemodels} with \Sec~\ref{sec:sima} and \Sec~\ref{sec:mp2} \\
Sim-B & Simplified model with factorized response function and resonance treatment~B (step function approximation) & Single integration, 23 IT, $\sim~(2-3)\times T_\mathrm{Sim-C}$  & \Sec~\ref{sec:simplemodels} with \Sec~\ref{sec:simb} and \Sec~\ref{sec:mp2}\\
Sim-C & Simplified model with factorized response function and resonance treatment~B; simplified multi-pion production & Single integration, 10 IT, $T_\mathrm{Sim-C}$  & \Sec~\ref{sec:simplemodels} with \Sec~\ref{sec:simb} and \Sec~\ref{sec:mp1} \\
\hline
\end{tabular}
\end{small}
\end{center}
\mycaption{\label{tab:models}Considered models for the comparison of approaches. The (computational) complexity decreases from the top to the bottom. The time needed for the computation of the photohadronics in model Sim-C is given by $T_\mathrm{Sim-C}$. The comparison of the computation times is done for power law spectra. For the computation with SOPHIA we used 100000 trials per proton bin.}
\end{table}

The models considered in this subsection are listed in \Tab~\ref{tab:models}, where the (computational) complexity decreases from the top to the bottom. ``SOPHIA'' represents the output of the SOPHIA software, computed for our benchmark spectra from \App~\ref{app:benchmarks}. The computation with SOPHIA is described in \App~\ref{app:SOPHIA}. The model ``BW'' corresponds to the basic physics of SOPHIA as described in \Sec~\ref{sec:photo}, including double integration for the direct production and multi-pion production from \Sec~\ref{sec:mp2}. In this case, we use Eqs.~(\ref{equ:prodmaster}) and~(\ref{equ:response}) for the secondary production using the two dimensional response function including all resonances with full Breit-Wigner forms and the direct production as described in \App~\ref{app:directkin}. Compared to SOPHIA, the kinematics is considerably simplified. The multi-pion production is taken from \Sec~\ref{sec:mp2} close to SOPHIA.
 ``Sim-A'' and ``Sim-B'' are the simplified models from \Sec~\ref{sec:simplemodels}, using the factorized response function introduced in \Sec~\ref{sec:rsimple}. Whereas Sim-A treats all interaction types with the resonances explicitely, Sim-B defines an interaction type for the $\Delta(1232)$-resonance, and one for the higher resonances. Therefore, Sim-B uses considerably less interaction types. Note that we have obtained Sim-A and Sim-B by condensing the information from BW stepwise. Both models use the multi-pion production from \Sec~\ref{sec:mp2}, whereas Sim-C is a simplified version of Sim-B including the multi-pion production in \Sec~\ref{sec:mp1}.

\begin{figure}[ht]
\begin{center}
\includegraphics[width=0.85\textwidth]{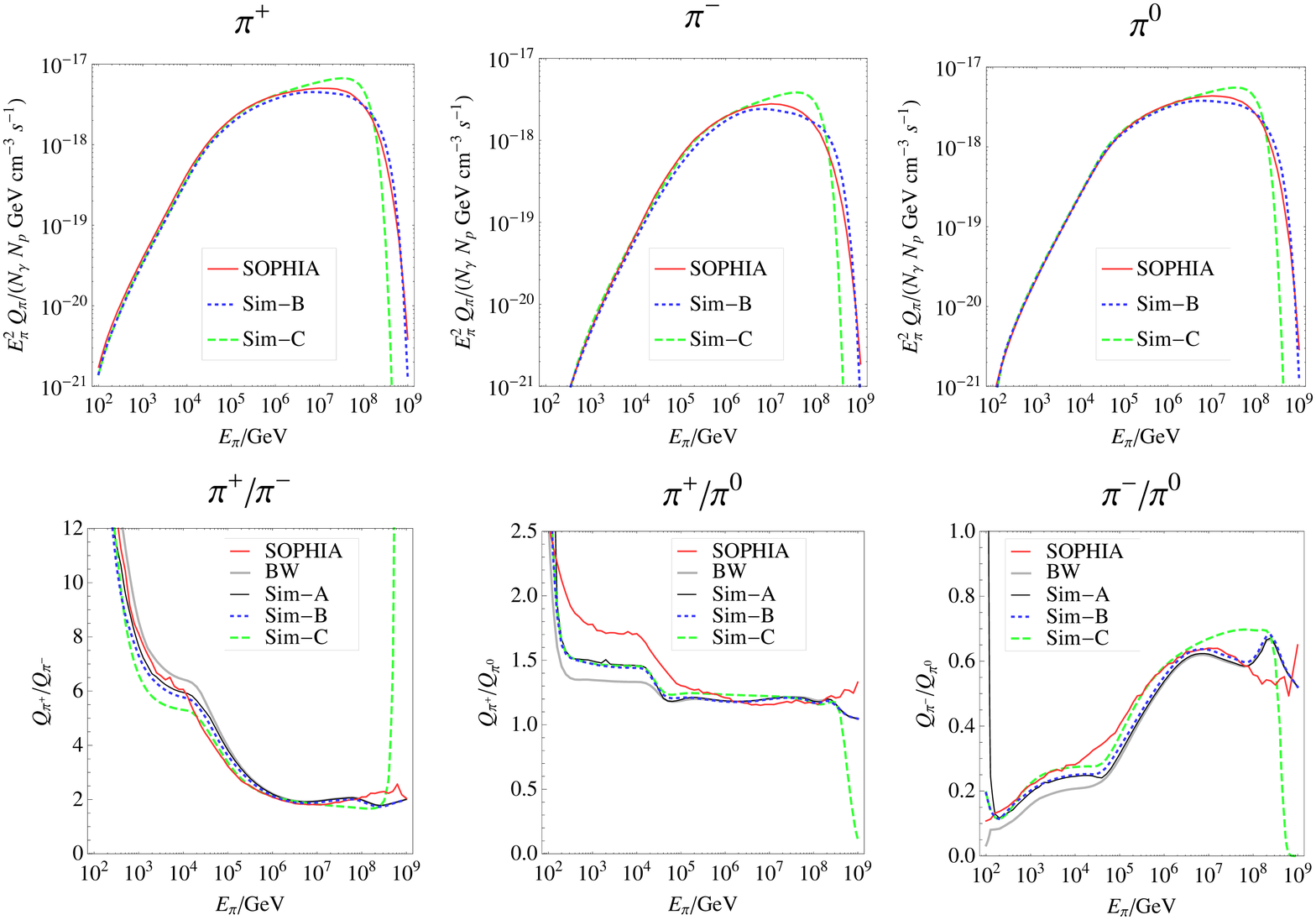}
\end{center}
\vspace*{-0.5cm}
\mycaption{\label{fig:comparisonpi} Comparison of pion spectra (upper panel) and pion ratios (lower panel) for the different models from \Tab~\ref{tab:models} for the GRB benchmark (see \App~\ref{app:benchmarks}). Since the pion spectra for model BW, Sim-A and Sim-B can not be distinguished, we plot only the results of SOPHIA, model Sim-B and model Sim-C in the upper row.}
\end{figure}
\begin{figure}[ht]
\begin{center}
\includegraphics[width=0.85\textwidth]{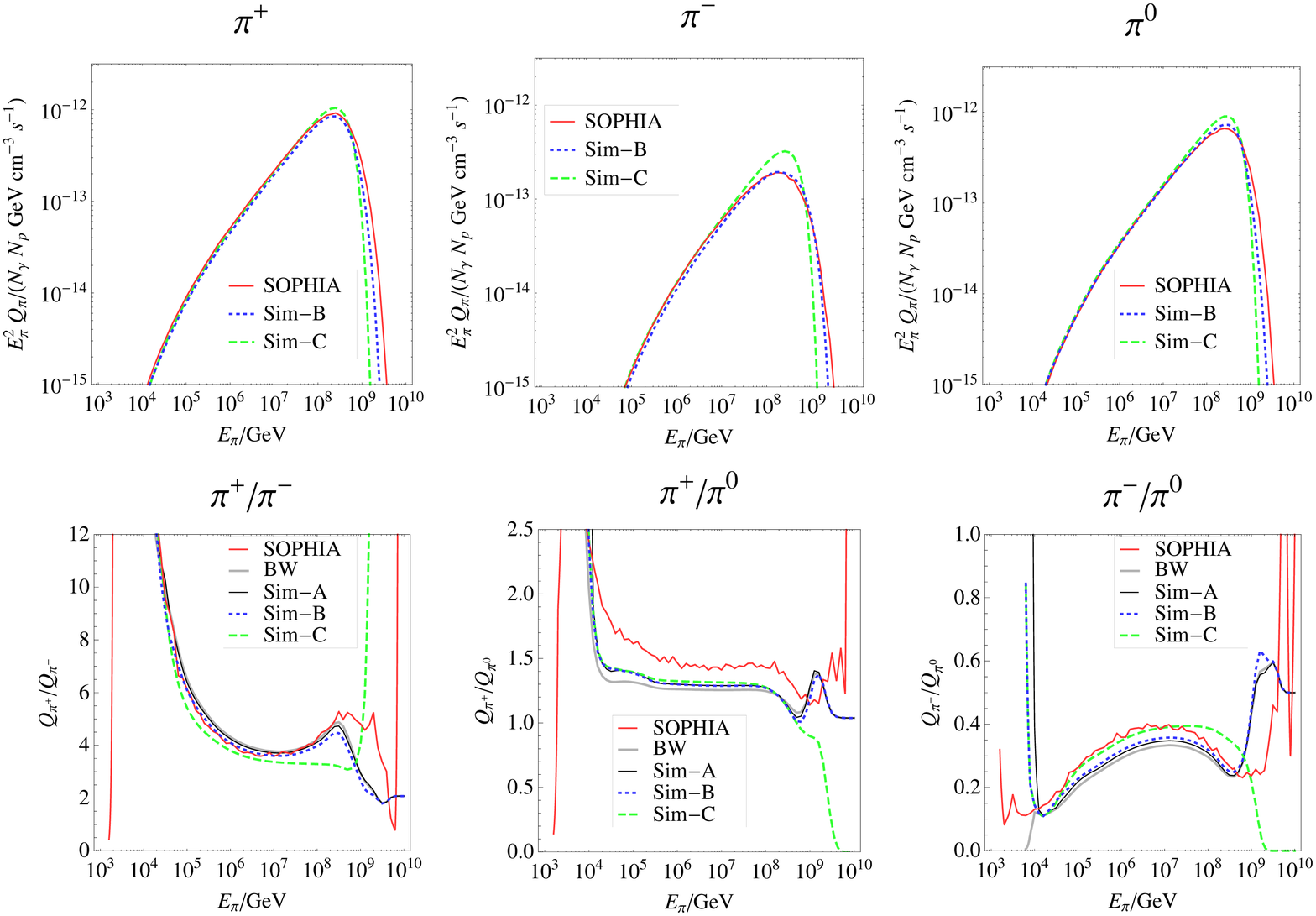}
\end{center}
\vspace*{-0.5cm}
\mycaption{\label{fig:comparisonpiagn} Comparison of pion spectra (upper panel) and ratios (lower panel) for the different models from \Tab~\ref{tab:models} for the AGN benchmark (see \App~\ref{app:benchmarks}). Since the pion spectra for model BW, Sim-A and Sim-B can not be distinguished, we plot only the results of SOPHIA, model Sim-B and model Sim-C in the upper row.}
\end{figure}
\begin{figure}[ht]
\begin{center}
\includegraphics[width=0.85\textwidth]{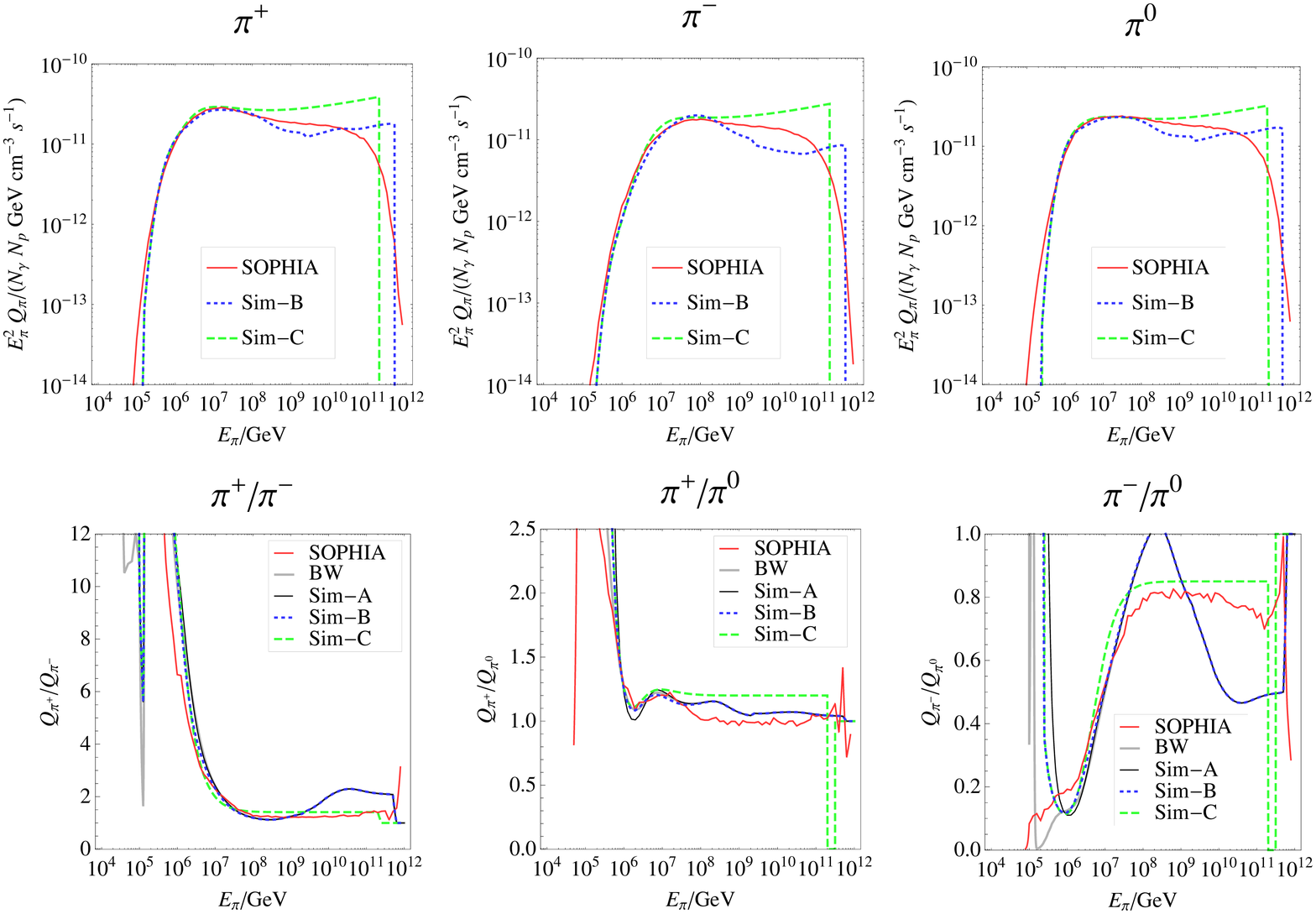}
\end{center}
\vspace*{-0.5cm}
\mycaption{\label{fig:comparisonpibb} Comparison of pion spectra (upper panel) and ratios (lower panel) for the different models from \Tab~\ref{tab:models} for the BB benchmark (see \App~\ref{app:benchmarks}). Since the pion spectra for model BW, Sim-A and Sim-B can not be distinguished, we plot only the results of SOPHIA, model Sim-B and model Sim-C in the upper row.}
\end{figure}

We compare the pion spectra produced by these different models in \figu{comparisonpi} (GRB benchmark), \figu{comparisonpiagn} (AGN benchmark) and \figu{comparisonpibb} (Black body (BB) benchmark). In these figures, the upper rows show the pion spectra for $\pi^+$, $\pi^-$, and $\pi^0$ explicitly, whereas the lower rows show the pion ratios $\pi^+/\pi^-$, $\pi^+/\pi^0$, and $\pi^-/\pi^0$. Since the pion spectra for model BW, Sim-A and Sim-B are so close to each other that they can not be distinguished, we plot only the results of SOPHIA, model Sim-B and model Sim-C in the upper rows of Figs.~\ref{fig:comparisonpi} to~\ref{fig:comparisonpibb}. Whereas the charged pions lead to neutrino production, the neutral pions lead to photons. Therefore, the ratio $\pi^+/\pi^0$ determines, to leading order, the ratio between neutrinos and photons, which also often enters the computation of neutrino flux limits. On the other hand, the ratio $\pi^+/\pi^-$ affects the (electron) neutrino-antineutrino ratio. The ratio $\pi^-/\pi^0$ is shown for completeness. Note that the normalization of the different spectra is not chosen arbitrarily, but consistently to be able to compare the spectra directly; \cf, \App~\ref{app:SOPHIA}.

As the most important fact, as it can be read off from the upper rows of the three figures, all the spectra of our simplified models (apart from Sim-C, maybe) match the output from SOPHIA very well, both normalization and spectral shape. At high energies, however, where we imposed a sharp spectral cutoff, the spectra from SOPHIA are smeared out because of the more refined kinematics treatment, which can best be seen in \figu{comparisonpibb} for the BB benchmark where we have a sharp spectral cutoff in the proton spectrum. This difference is unavoidable, and the price to pay for an efficient simplified model. Nevertheless, note that in more realistic spectra, or spectra averaged over different sources, such sharp features in the spectra may not be present. Although it seems that there are hardly any differences between SOPHIA and our models at lower energies, one can read off from the pion ratio plots in the lower rows (on a linear vertical scale) that there are small deviations. Whereas the differences at very low and high energies, where the spectra rapidly break off, are not very surprising, there are some differences coming from the different kinematics treatment. For example, SOPHIA actually produces even about 20\% more $\pi^+$ than $\pi^0$ in the lower energy range for the GRB benchmark and the middle energy range for the AGN benchmark (\cf, middle lower panels). We have checked that this difference neither comes from the resonance or direct production treatment, nor from the relative contribution of both processes. Instead, for some processes, our kinematics is a bit over-simplified, since for these a considerable amount of pions is (in SOPHIA) reconstructed at lower energies. Another example for an obvious difference is the high energy $\pi^+$/$\pi^-$ and $\pi^+$/$\pi^0$ difference in the lower panel of \figu{comparisonpibb} for the BB benchmark, the most challenging one for the treatment of multi-pion production. The discrepancy is a result of the simplified kinematics treatment of taking the same $\chi$-values for $\pi^+$, $\pi^0$ and $\pi^-$. This mainly effects the $\pi^-$ spectra because they have, in comparison to $\pi^+$ and $\pi^0$, slightly different kinematics in SOPHIA. In the upper panel in \figu{comparisonpibb} a double hump structure can be seen which follows from the kinematics treatment of multi-pion production that one part of the pions is reconstructed at lower energies (small $\chi$) and the other at higher energies (large $\chi$). If one averages over larger energy scales (such as in diffuse fluxes), such kinematics effects average out.

As far as the comparison among our simplified models is concerned, the differences are small compared to the effects of kinematics discussed above. In fact, model ``BW'', which was originally designed as the most accurate reproduction of SOPHIA, produces the results farthest off from SOPHIA, especially for the lower half of the energy range. The reason may be that the errors introduced
by the approximations in Sim-A and Sim-B partly compensate the errors from the simplified multi-pion production and direct channels. The model Sim-A was obtained from BW by assuming that all resonances are strongly peaked, whereas Sim-B was derived from this assumption by collecting the properties of the higher resonances into one interaction type. In the comparison of model Sim-B to Sim-C the differences between the kinematics for multi-pion production from \Sec~\ref{sec:mp2} and the simplified one from \Sec~\ref{sec:mp1} can nicely be seen. For the GRB (see \figu{comparisonpi}) and the AGN benchmark (see \figu{comparisonpiagn}) this mainly effects the high energy region whereas in the BB case (see \figu{comparisonpibb}) most of the energy range is affected. 
Since the computation time for Sim-B, for which the results do not differ significantly from model Sim-A, is only about a factor of 2-3 longer and the high energy treatment is way more accurate than for Sim-C (especially close to the peaks), we focus in the following on model Sim-B. As one can see in Figs.~\ref{fig:comparisonpi} to~\ref{fig:comparisonpibb}, Sim-B is most accurate for power laws which are our main interest. Compared to SOPHIA, we gain a factor of about 1000 (if implemented in C, depending on the integration method) in computation time for 100000 trials per proton bin in SOPHIA (as we use for the GRB benchmark). The Sim-B spectra do not have any small wiggles, because the computation is exact. However, note that the complexity of Sim-B increases with the number of interaction types, whereas the complexity of SOPHIA increases with the number of trials (and required smoothness of the functions).

\subsection{Neutrino spectra}

In this section, we first review the production of neutrinos from pion and subsequent muon decays,
as well as kaon decays.
For the sake of completeness, we include the neutrinos from neutron decays, where the
neutrons are produced by the photohadronic interactions. Then we compare our results to SOPHIA. We focus on model Sim~B from the previous section only.

The neutrinos are mostly produced in the following two decay chains:
\begin{eqnarray}
\pi^+ & \rightarrow & \mu^+ + \nu_\mu \, ,\nonumber \\
& & \mu^+ \rightarrow e^+ + \nu_e + \bar{\nu}_\mu \, , \label{equ:piplusdec} \\
\pi^- & \rightarrow & \mu^- + \bar\nu_\mu \, , \nonumber \\
& & \mu^- \rightarrow e^- + \bar\nu_e + \nu_\mu \, . \label{equ:piminusdec}
\end{eqnarray}
For the energy spectrum from weak decays, we follow \citet{Lipari:2007su} (\Sec~IV). In this case,  the decays are
described by \equ{prod}. The functions $d n_{a \rightarrow b}^{\mathrm{IT}}/dE_b$
simplify, in a frame where the parent $a$ is ultra-relativistic, to
$d n_{a \rightarrow b}^{\mathrm{IT}}/dE_b = 1/E_a \, F_{a \rightarrow b} \left( E_b/E_a \right) $. The functions  $F_{a \rightarrow b}$ include the measured branching ratios in the possible
final states and are given in \citet{Lipari:2007su} (\Sec~IV). Note that $\pi^+$ and $\pi^-$ are initially produced in different ratios and produce muons with different helicities, described by the scaling functions with $r_\pi=(m_\mu/m_\pi)^2$:
\begin{eqnarray}
 F_{\pi^+\rightarrow\mu_R^+}(x)=&F_{\pi^-\rightarrow\mu_L^-}(x)&=\frac{r_\pi(1-x)}{(1-r_\pi)^2x}\Theta(x-r_\pi)\\
 F_{\pi^+\rightarrow\mu_L^+}(x)=&F_{\pi^-\rightarrow\mu_R^-}(x)&=\frac{x-r_\pi}{(1-r_\pi)^2x}\Theta(x-r_\pi).
\end{eqnarray}
The muons decay further in a helicity-dependent way:
\begin{eqnarray}
 F_{\mu^+\rightarrow\bar{\nu}_\mu}(x,h)= &F_{\mu^-\rightarrow\nu_\mu}(x,-h)&=\left(\frac{5}{3}-3x^2+\frac{4x^3}{3}\right)+h\left(-\frac{1}{3}+3x^2-\frac{8x^3}{3}\right)\\
  F_{\mu^+\rightarrow\nu_e}(x,h)= &F_{\mu^-\rightarrow\bar{\nu}_e}(x,-h)&=\left(2-6 x^2+4 x^3\right)+h\left(2-12 x+18x^2-8 x^3\right)
\end{eqnarray}
with $h=1$ for right-handed and $h=-1$ for left-handed muons. It is therefore mandatory to distinguish four muon states $\mu_L^+$, $\mu_R^+$, $\mu_L^-$, and $\mu_R^-$ as final states in order to account for the impact of muon polarization. The decay rates $\Gamma_{a \rightarrow b}^{\mathrm{IT}}= \Gamma$ in \equ{prod} (there is only one interaction type, which is decay) are just the inverse lifetimes $\Gamma = \tau^{-1} = \tau_0^{-1} \, m_a \, E_a^{-1}$, where $\tau_0$ is the rest frame lifetime.

For kaons, the leading decay mode into muon and neutrino is treated in the same way as in \citet{Lipari:2007su} for the pion decays, \ie, with $m_\pi \rightarrow m_K$. The branching ratio for this channel is about 63.5\%.
The second-most-important decay mode is $K^\pm \rightarrow \pi^\pm +  \pi^0$ (20.7\%).
The other decay modes account for 16\%, no more than about 5\% each.
Since interesting effects can only be expected in the energy range with the most energetic neutrinos, we only use the direct decays from the leading mode.

For protons accelerated in the jet, neutrons are produced by photohadronic interactions as described in \App~\ref{app:others}. Assuming that the neutrons escape from the acceleration region before they interact again, an additional neutrino flux from neutron decays is obtained. In this section, we show this neutrino flux separately. The beta decay describes the decay of the neutron into a proton, an electron and an electron anti-neutrino. In the ultra-relativistic case, the mean fraction of the neutron energy going into the neutrino is $\chi\approx5.1\times10^{-4}$, see \citep{Lipari:2007su}. The neutrino spectrum is therefore obtained from the following equation:
\begin{equation}
 Q_{\bar{\nu}_e}(E_\nu)=\frac{1}{\chi} Q_n\left(\frac{E_\nu}{\chi}\right)
\label{equ:ndecay}
\end{equation}
with $Q_n$ calculated from \equ{reinj} (\App~\ref{app:others}).

\begin{figure}[t!]
\begin{center}
\includegraphics[width=\textwidth]{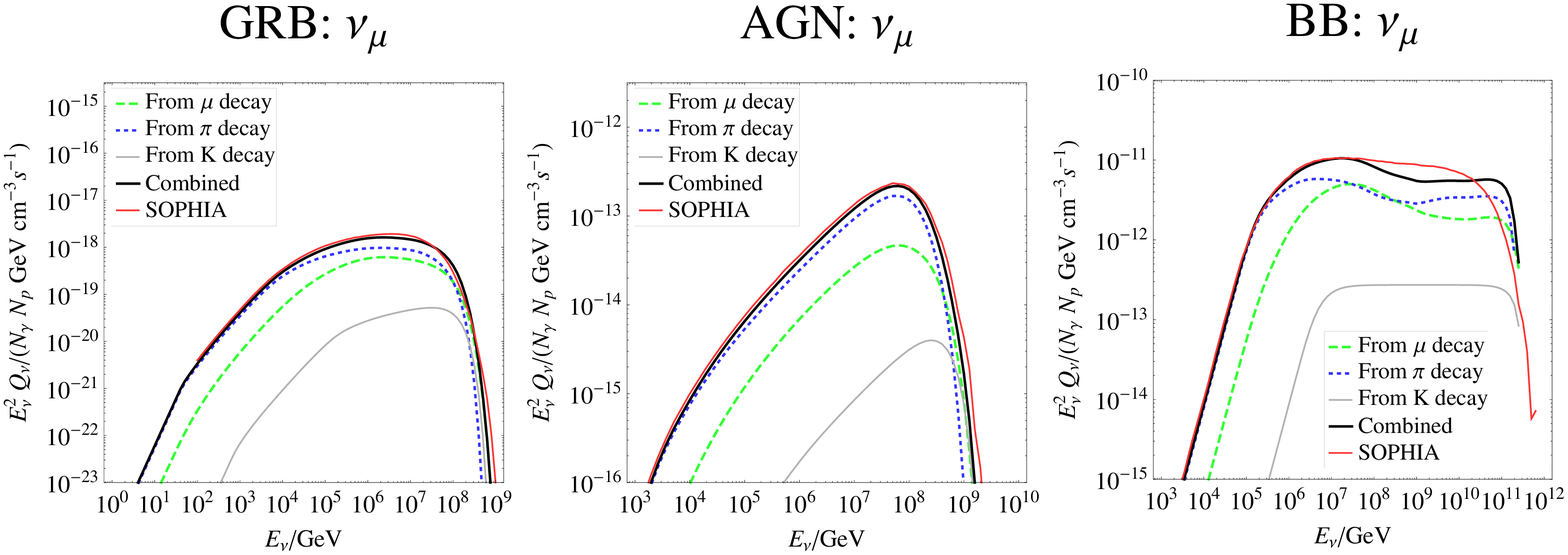}
\end{center}
\mycaption{\label{fig:comparisonnu} Comparison of $\nu_\mu$ spectra from the decays of different parents, as denoted in the labels. The left panel is for the GRB benchmark, the middle one for the AGN benchmark and the right one for the BB benchmark (see \App~\ref{app:benchmarks}).}
\end{figure}

We show the $\nu_\mu$ neutrino spectra obtained from Sim-B and SOPHIA in \figu{comparisonnu}, where we also show the different contributions from different decay modes separately. Obviously, the SOPHIA and our combined spectra match very well, apart from the already discussed difference in the kinematics leading to some averaging in SOPHIA for large energies.
Since the production of $\pi^+$ dominates for initial protons, $\nu_\mu$ in \figu{comparisonnu} are most abundantly produced from pion decay; \cf, \equ{piplusdec}. However, the $\nu_\mu$ from muon decays, coming from the $\pi^-$ decay chain -- \cf, \equ{piminusdec} -- are found at slightly higher energies and dominate for every high energies in the spectrum. For $\bar\nu_\mu$, the situation is exactly the opposite, but the final spectra look very similar. For very high energies, the SOPHIA spectrum is slightly higher than what one would expect, because other decay modes (such as from neutral kaons) contribute, which we have not considered.
Without synchrotron cooling, the contribution from kaon decays is, however, small.

\begin{figure}[t!]
\begin{center}
\includegraphics[width=\textwidth]{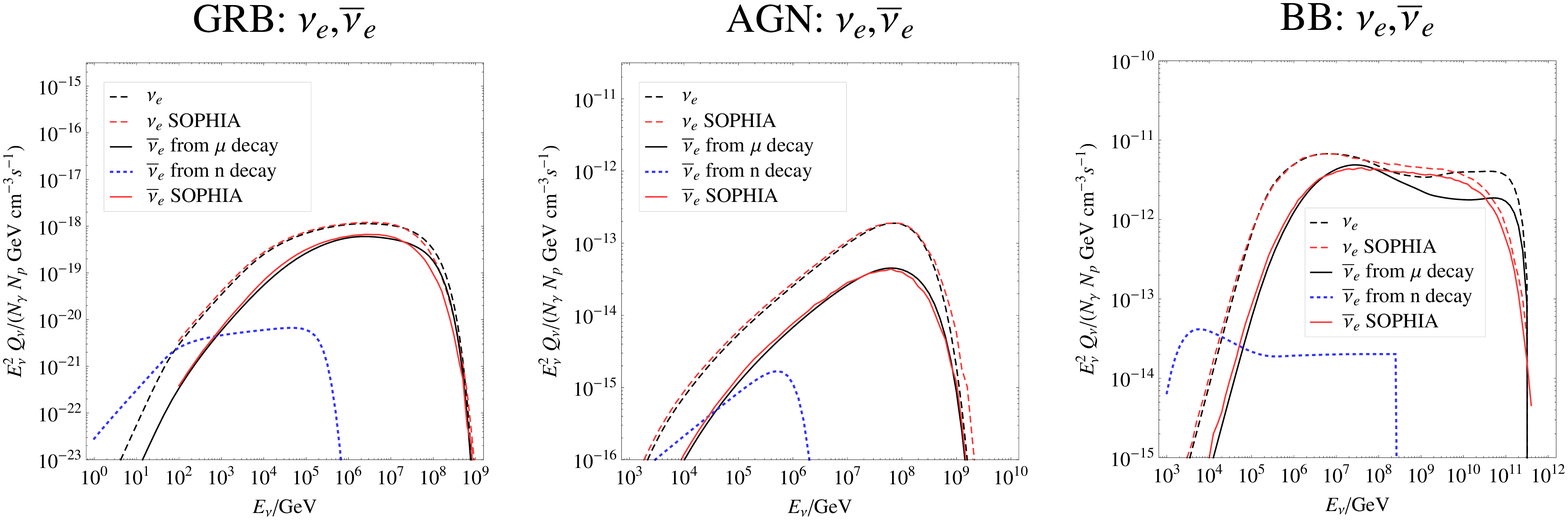}
\end{center}
\mycaption{\label{fig:comparisonenu} Comparison of the $\nu_e$ (dashed) and $\bar \nu_e$ (solid) spectra from the decays of different parents, as denoted in the labels. The left panel is for the GRB benchmark, the middle one for the AGN benchmark and the right one for the BB benchmark (see \App~\ref{app:benchmarks}).}
\end{figure}

In \figu{comparisonenu}, we show the $\nu_e$ (dashed) and $\bar \nu_e$ (solid) fluxes for our benchmarks. Obviously, the $\nu_e$, coming from the $\pi^+$ decay chain in \equ{piplusdec}, dominate over the $\bar \nu_e$. However, if the neutrons produced by the photohadronic interactions escape from the source and then decay, they will lead to an additional neutrino flux shown by the thin gray curves (not included in the total $\bar\nu_e$ curves). Especially at very low energies, the $\bar \nu_e$ flux then dominates.

\subsection{Flavor and neutrino-antineutrino ratios of the neutrinos}

In this section, we discuss the electron to muon neutrino flavor ratio (the ratio between the electron and muon neutrino fluxes) and the neutrino-antineutrino ratios at the source. The flavor composition at the source is primarily characterized by the flux ratio  $(\bar \nu_e+ \nu_e)/(\bar \nu_\mu + \nu_\mu)$, since almost no $\nu_\tau$ (or $\bar \nu_\tau$) are expected to be produced at the source. Because neutrino telescopes can, in muon tracks or showers, not distinguish neutrinos from antineutrinos, this ratio is representative for the detection as well. From Eqs.~(\ref{equ:piplusdec}) and~(\ref{equ:piminusdec}), we can read off that this flavor ratio should be about $1/2$ without kinematical effects, which we call the ``standard assumption''.  At the Earth, the three neutrino flavors then almost equally mix (in the ratio 1:1:1) through neutrino flavor mixing~\citep{Learned:1994wg}.

\begin{figure}[t!]
\begin{center}
\includegraphics[width=\textwidth]{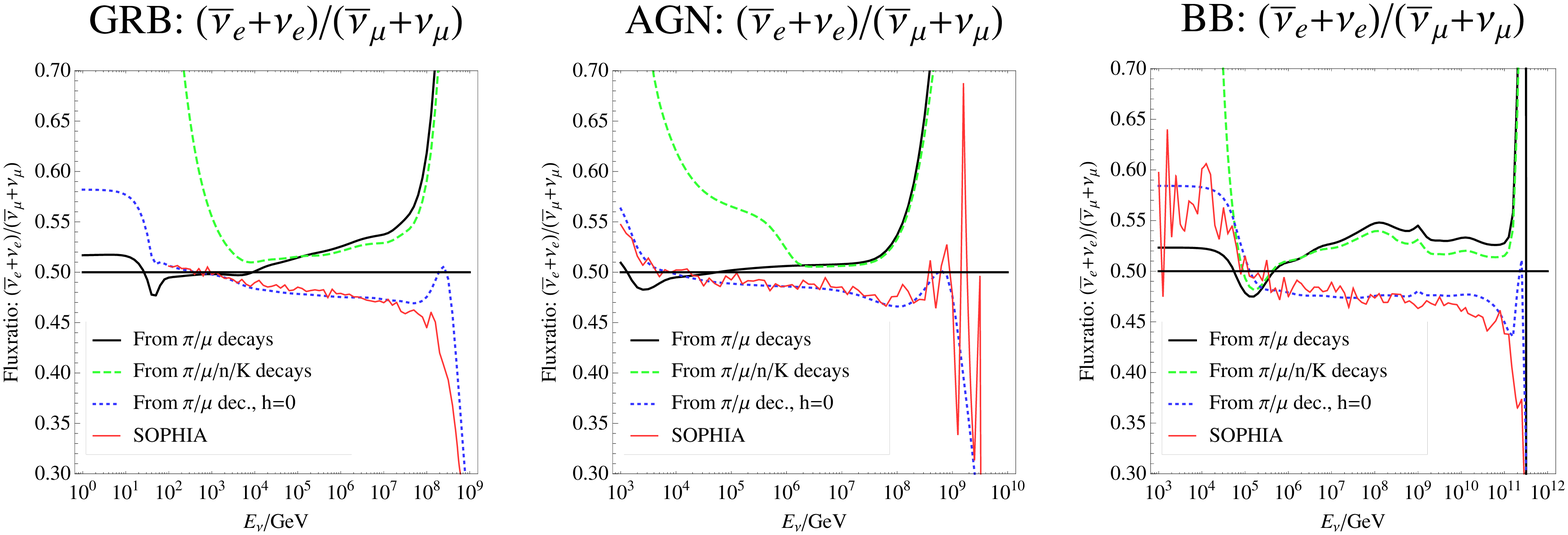}
\end{center}
\mycaption{\label{fig:flavor} Comparison of total electron to muon neutrino flavor ratio at the source for the following curves as given by the labels: Neutrinos from pion/muon decays, neutrinos from these including neutron and kaon decays, neutrinos from Sim-B (pion/muon decays) for without taking into account the spin state of the final muon ($h=0$), and SOPHIA output. The horizontal lines mark the ``standard'' assumption for a flavor ratio electron to muon neutrinos 1:2. The left panel is for the GRB benchmark, the middle one for the AGN benchmark and the right one for the BB benchmark (see \App~\ref{app:benchmarks}).}
\end{figure}

We show in \figu{flavor} the flavor ratio at the source for the GRB (left panel), AGN (middle panel) and BB (right panel) benchmark.
The standard assumption 1/2 is marked by the horizontal lines. Our curves from Sim-B (thick solid, neutrinos from pion and muon decays only) bend upward from this standard assumption, whereas the SOPHIA curves bend downward for large energies. This difference can be explained by a different implementation of the weak decays: if we do not take into account the spin state of the final muon (dotted curves $h=0$), we can reproduce the SOPHIA results almost exactly with Sim-B. In fact, the effect of the helicity is larger than the details of the interaction model. The dashed curves include the effect of neutron decays (low energies) and kaon decays (high energies) into Sim-B. Especially for low energies, where $\bar \nu_e$ are abundantly produced by neutron decays, the curves deviate from the thick ones. This effect is strongest for the AGN benchmark, for which the standard assumption  1/2 only approximately holds in a relatively narrow energy window.
Note that the dashed curve in the left panel and all the other results for the GRB benchmark exactly match \citet{Lipari:2007su}, where the weak decays are discussed in detail.

\begin{figure}[t!]
\begin{center}
\includegraphics[width=\textwidth]{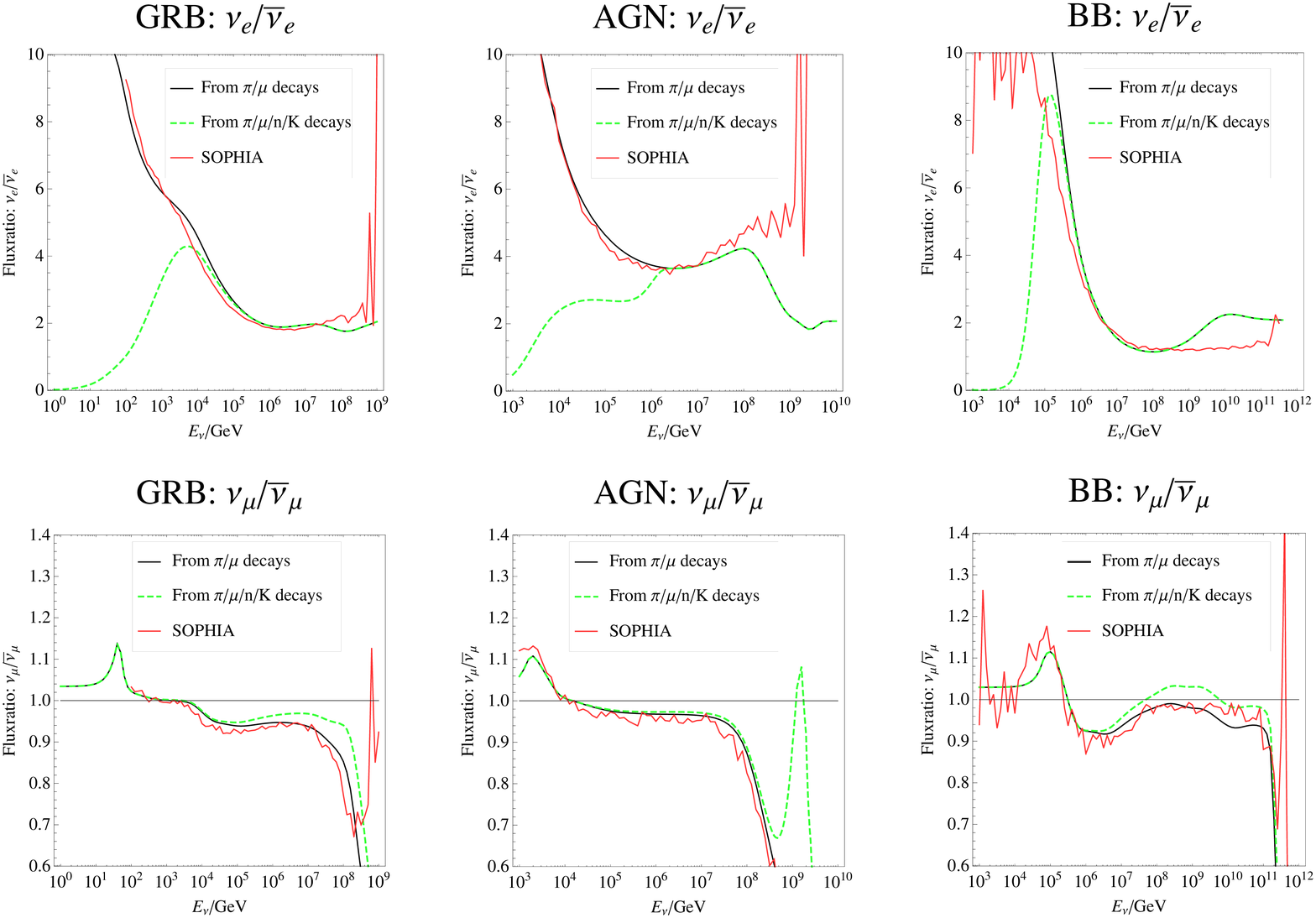}
\end{center}
\mycaption{\label{fig:neutrinoantineutrino} Comparison of the neutrino-antineutrino ratios at the source for electron neutrinos (left panels) and muon neutrinos (right panels) for the following curves given by the labels: Neutrinos from pion/muon decays, neutrinos from these including neutron and kaon decays, and SOPHIA output. The horizontal lines in the lower panels mark the ``standard'' assumption for a flavor ratio muon neutrinos to antineutrinos 1:1. The left panels are for the GRB benchmark, the middle ones for the AGN benchmark and the right ones for the BB benchmark  (see \App~\ref{app:benchmarks}).}
\end{figure}

If the Glashow resonance process $\bar{\nu}_e + e^- \to W^- \to \text{anything}$ at around $6.3 \, \text{PeV}$ can be observed in a neutrino telescope, the neutrino-antineutrino ratios at the source may be relevant as well (all flavors at the source contribute to the $\bar\nu_e$ flux at the Earth through flavor mixing)~\citep{Learned:1994wg,Anchordoqui:2004eb, Bhattacharjee:2005nh}. The neutrino-antineutrino ratio may be relevant to distinguish $p\gamma$ interactions at the source, for which mostly $\pi^+$ are produced, from $pp$ interactions at the source, for which $\pi^+$ and $\pi^-$ are produced in almost equal ratios. Therefore, we show in \figu{neutrinoantineutrino} the neutrino-antineutrino ratios at the source for electron neutrinos (left panels) and muon neutrinos (right panels). The electron neutrino-antineutrino ratios in the left panels depend on the ratio of $\pi^+$ and $\pi^-$ produced, see Eqs.~(\ref{equ:piplusdec}) and~(\ref{equ:piminusdec}). Our result matches SOPHIA very well, especially in the important energy range from the peak of the spectrum two decades down, apart from the discrepancy for high energies for the BB benchmark (upper right panel) which we discussed already as it is coming from the $\pi^+/\pi^-$ ratio.
The deviation between SOPHIA and Sim-B can be up to 30\%. After flavor mixing, the correction to the electron neutrino-antineutrino ratio at the Earth is at the level of 10\%, much smaller than the effect on the flavor mix expected from $pp$ interactions. The muon neutrino-antineutrino ratios in the right panels do not depend on the ratio of $\pi^+$ and $\pi^-$ produced, as in every pion decay the same number of muon neutrinos and antineutrinos is produced; see Eqs.~(\ref{equ:piplusdec}) and~(\ref{equ:piminusdec}). In this case, our results match SOPHIA and the standard prediction very well, and the effects of neutron and kaon decays are small in the absence of cooling.

\section{Summary and conclusions}
\label{sec:summary}

We have discussed simplified models for photo-meson production in cosmic accelerators. The main purpose of this simplification has been the definition of a photohadronic interaction model useful for efficient modern time-dependent AGN and GRB simulations, and for large-scale parameter studies, such as of neutrino flavor ratios. The major requirements have been listed in the introduction. For example, the secondaries (pions, kaons) are not to be integrated out, since their synchrotron cooling affects the neutrino flavor ratios.

We have first re-phrased the problem in terms of a two-dimensional ``response function'' to be folded with arbitrary photon and proton input spectra in order to compute the secondary fluxes.  The key idea for our simplified models has been the factorization of this two-dimensional response function, which has allowed to eliminate one of the integrations. In order to include kinematics as good as possible, we have then defined a discrete number of different interaction types with different characteristics based on the underlying physics of SOPHIA. The kinematics of more complicated interactions, such as direct production or  multi-pion production, has been simulated by the introduction of multiple interaction types for each production channel. In a step-by-step fashion we have simplified then the resonance treatment in order to arrive at our simplified model Sim-B. It allows for the computation of pion spectra with only one integral, summed over about ten interaction types, and can be easily adopted from our description. The extendibility of this approach has been demonstrated by showing how $K^+$ fluxes can be added, once a suitable parameterization is found.

Similarly, the response function can be easily changed if new data are provided, new processes can be included, or systematics on the particle physics can be added. Of course, some effort has to be spent to find a suitable parameterization for each process.

We have demonstrated that our results match the output of SOPHIA sufficiently well. However, there are some differences due to the more refined kinematics treatment of SOPHIA, which effectively corresponds to one additional integration. For example, for very narrow spectral features, such as rapid cutoffs, the spectra are naturally more smeared out by SOPHIA. This is especially the case for high energetic interactions where multi-pion production is dominant, as can be seen in the BB (black body) benchmark. However, our approach is much simpler in the sense that the interaction rate and the folding with the proton spectra is automatically taken into account (\cf, \App~\ref{app:SOPHIA}). In addition, we have included the spin state of the final muon in the pion decays, as described in \citet{Lipari:2007su}, not included in SOPHIA, which leads to differences in the neutrino flavor ratios: in fact, the electron to muon neutrino flavor ratio at the source is typically larger than 0.5, instead of smaller, as predicted without taking into account the spin state. In particular, we obtain $\nu_e:\nu_\mu:\nu_\tau\simeq$ $1:1.85:0$ for the GRB benchmark, $1:1.96:0$ for the AGN benchmark, and up to $1:1.82:0$ for the BB benchmark close to the spectral peaks. This means that especially the AGN benchmark behaves as pion beam in spite of the helicity dependence of the muon decays, whereas the BB benchmark shows the strongest deviation.

Since our approach has allowed us to discuss the leading interaction types separately, we have also shown the differences to the $\Delta(1232)$-resonance approximation (see also \citet{2000NuPhS..80C0810M, Mucke:1999fh}). For example, we have shown how multi-pion production modifies the characteristic shape of the GRB neutrino spectrum expected from the resonance approximation (which is flat in $E^2$ times the flux in the intermediate energy range). In fact, all of the the resonances combined do not dominate in any significant part of the charged pion spectrum for our GRB, AGN and BB benchmarks. In addition, the $\Delta(1232)$-resonance approximation has rendered insufficient for the computation of the (electron) neutrino-antineutrino ratios at the source, because, by definition, no $\pi^-$ are produced. We have also demonstrated from our general response function in model Sim-B, that for any input proton or photon spectrum the $\pi^+/\pi^0$ ratio at the source is significantly larger than one, as opposed to 1/2 from the $\Delta(1232)$-resonance approximation. This implies that any neutrino flux based on the $\Delta(1232)$-resonance approximation and normalized to photon flux observations using the charged to neutral pion ratio is underestimated by at least a factor of $2.4$, where this factor is independent of the input spectra.

We conclude that our simplified model Sim-B allows for an efficient computation of pion and neutrino fluxes, including the necessary features for neutrino flux ratio discussion and the necessary efficiency for time-dependent simulations. It is sufficient for many purposes especially for power law photon fields, but, of course, it cannot replace a full Monte Carlo simulation including full kinematics if high precision fits of existing data are required. In particular, it may turn out to be useful for time-dependent simulations and extensive parameter space studies using power law spectra, including spectral breaks. However, in other cases, such as if the high energy interactions with photon spectra with sharp peaks are very important,  or there are anisotropic photon distributions, the Monte Carlo method may be the better choice.

\subsubsection*{Acknowledgments}

We would like to thank M. Maltoni, K. Mannheim, D. Meloni, and A. Reimer for useful discussions.
SH acknowledges support from the Studienstiftung des deutschen Volkes (German National Academic Foundation). FS acknowledges support from Deutsche Forschungsgemeinschaft through grant SP 1124/1.
WW would like to acknowledge support from the
Emmy Noether program of Deutsche Forschungsgemeinschaft,
contract WI 2639/2-1. This work has also been supported by the Research Training Group GRK1147 ``Theoretical astrophysics and particle physics'' of Deutsche Forschungsgemeinschaft.

\begin{appendix}

\section{Kinematics treatment of direct production}
\label{app:directkin}

\begin{figure}[ht]
\begin{center}
\includegraphics[width=0.7\textwidth]{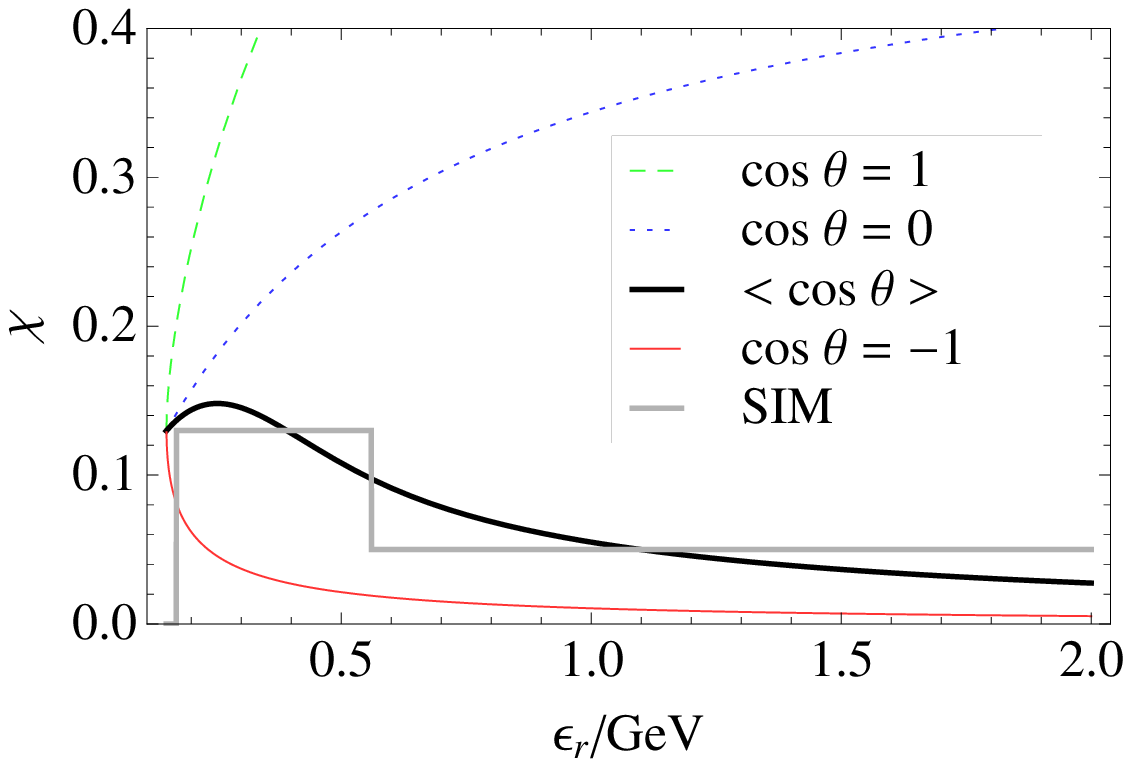}
\mycaption{\label{fig:kinematic} Energy fraction $\chi$ going into the pion as a function of $\epsilon_r$ for direct one pion production for different values of $\cos{\theta}$ and our simplified approach ``SIM''; see main text for details}
\end{center}
\end{figure}

 Here we follow the approach of \citet{Rachen:1996ph}. As mentioned in \Sec~\ref{sec:kinsec}, the direct production process is strongly backward peaked with respect to the produced pion. One possible approximation for the determination of $\chi$ is to assume that the average scattering angle is $180^\circ$. The energy fraction $\chi$ going into the pion as a function of $\epsilon_r$ is shown in \figu{kinematic} as the curve $\cos \theta =-1$ (\cf, Eq.~(\ref{eqn:1pion})). For comparison, the curves $\cos \theta =0$ and $\cos \theta =1$ are shown as well. For a more accurate representation of kinematics, we take the probability distribution of the Mandelstam variable $t$, as given in  \citet{Mucke:1999yb}, into account. For small $|t|$, it is given by:
\begin{equation}
\frac{dP}{dt}=b\frac{e^{b\,t(\cos{\theta_{\pi}})}}{e^{b\,t(-1)}-e^{b\,t(1)}}
\label{equ:tdis}
\end{equation}
 with $b\approx12\, \mathrm{GeV^{-2}}$.
For example, for the interaction type $\mathrm{T}_1$,  the Mandelstam variable $t$  is given by
\begin{equation}
 \label{equ:mandelt}
t(\cos{\theta_{\pi}})=m_p^2-2 \frac{s+m_p^2}{2\sqrt{s}}\frac{s+m_p^2-m_\pi^2}{2\sqrt{s}}-2\frac{s-m_p^2}{2\sqrt{s}}\sqrt{\left(\frac{s+m_p^2-m_\pi^2}{2\sqrt{s}}\right)^2-m_p^2} \, \cos{\theta_{\pi}}\,.
\end{equation}
Combining Eqs.~(\ref{equ:tdis}) and~(\ref{equ:mandelt}), we obtain the average scattering angle as a function of the center-of-mass energy. Inserting the result into Eq.~(\ref{eqn:1pion}), we obtain the average energy fraction $\chi$ going into the pion for direct one pion production. It is shown in \figu{kinematic} as curve ``$< \cos \theta >$'' as a function of $\epsilon_r$. Analogously we compute the mean $\chi$ for direct two pion production by combining \equ{tdis} with the variable $t$ for the considered process.

In the simplified model for direct production (\cf, \Sec~\ref{sec:directsimple}) we use the factorized response function (see \equ{split}) as for the simplified models of the other processes. Therefore we have to choose an energy independent, constant $\chi$. Since the range of $\chi$-values for direct production is wide, we divide the $\epsilon_r$-range into three sections (for each of the interaction types $\mathrm{T}_1$ and $\mathrm{T}_{2a}$), such that it reproduces the results of the non-simplified model for typical power law spectra for photons and protons in astrophysical sources, such as GRBs and AGNs. This approach corresponds to the thick gray  curve ``SIM'' in \figu{kinematic}, where only the lower two interaction types are visible in the plotted energy range.
Obviously, it is a good step function approximation to the curve ``$< \cos \theta >$''. The different sections of this step function correspond to our interaction types ``L'', ``M'', and ``H''.

\section{Cooling and escape of primaries, re-injection}
\label{app:others}

A related issue to the secondary particle production is the cooling or escape of the primaries due to the interaction process. We do not focus on the cooling or escape timescales in this study, but, for the sake of completeness, we demonstrate how they can be computed from the quantities presented for our simplified models in \Sec~\ref{sec:simplemodels}.  If the primaries lose energy in an interaction, such as protons or neutrons in pion photoproduction $p + \gamma \rightarrow p + \pi^0$, this process can be interpreted as cooling, whereas if the primaries disappear, such as protons in $p + \gamma \rightarrow n + \pi^+$, it can be interpreted as escape.
In the latter case, neutrons are re-injected into the system, which we will discuss below.
 The cooling rate $t_{\mathrm{cool}}^{-1}(E_p)=-E_p^{-1} dE_p/dt$ or escape rate $t_{\mathrm{esc}}^{-1}=-N_p^{-1} dN_p/dt$ for the species $p$ (proton or neutron) due to the photohadronic interactions is, for constant $K^{\mathrm{IT}}$, given by
\begin{equation}
t_{\mathrm{cool}}^{-1}(E_p) = \sum_{\mathrm{IT}} \Gamma_{p \rightarrow p}^{\mathrm{IT}}(E_p) \,  K^{\mathrm{IT}} \, , \quad t_{\mathrm{esc}}^{-1}(E_p) = \sum_{\mathrm{IT}, \, p' \neq p} \Gamma^{\mathrm{IT}}_{p \rightarrow p'}(E_p) \, .
\label{equ:cool}
\end{equation}
in terms of the quantities in \equ{prod}.
Here $ K^{\mathrm{IT}} \cdot E_p$ is the loss of energy per interaction; therefore, $ K^{\mathrm{IT}}$ is often called ``inelasticity''. Note that if it is a function of the kinematical variables, such as the center of mass  energy, it has to be folded into the calculation of the interaction rate in $t_{\mathrm{cool}}^{-1}$. However, in \Sec~\ref{sec:simplemodels}, we have constructed the interaction types such that $ K^{\mathrm{IT}}$ is a constant.  For photo-pion production, the inelasticity can be related to the $\chi_{a \rightarrow b}^{\mathrm{IT}}$ in \equ{nabsimple} by
\begin{equation}
K^{\mathrm{IT}} = \sum\limits_{b \neq p}  \chi_{p \rightarrow b}^{\mathrm{IT}} \,   M_b^{\mathrm{IT}} \, ,
\label{equ:kchi}
\end{equation}
\ie, the energy loss of the nucleon equals the energy deposited in all interaction products (other than the initial nucleon). Note that the classification as cooling or escape also depends on if protons and neutrons are distinguished in the final state. In this section, we distinguish protons and neutrons.
In addition, note that in astrophysical objects there may be other sources of cooling and escape to be taken into account, such as synchrotron cooling or escape from the production region.

The quantity needed for the computation of \equ{cool} is the interaction rate in \equ{pgamma2}. Comparing  \equ{pgamma2} with \equ{split}, we find for our simplified models that the interaction rate for interaction type IT of the initial nucleon $p$ is
\begin{equation}
\Gamma^{\mathrm{IT}}(E_p) = \int\limits_{\frac{\epsilon_{\mathrm{th}} m_p}{2 E_p}}^{\infty} d \varepsilon \, n_\gamma(\varepsilon) \,
f^{\mathrm{IT}}\left( \frac{E_p \varepsilon}{m_p} \right) \, ,
\label{equ:irates}
\end{equation}
\ie, conveniently parameterized in terms of our simplified response function.
Then the cooling and escape rates in \equ{cool} can be written in terms of the initial $M_{p}$ or different nucleon $M_{p'}$ multiplicity ($M_p+M_{p'}=1$):
\begin{equation}
t_{\mathrm{cool}}^{-1}(E_p) = \sum_{\mathrm{IT}} M_p^{\mathrm{IT}} \, \Gamma^{\mathrm{IT}}(E_p)  \,  K^{\mathrm{IT}} \, , \quad t_{\mathrm{esc}}^{-1}(E_p) = \sum_{\mathrm{IT}, \, p' \neq p} M_{p'}^{\mathrm{IT}} \, \Gamma^{\mathrm{IT}}(E_p) \, .
\label{equ:cool2}
\end{equation}
The nucleon multiplicities are for the resonances in model~A given in \Tab~\ref{tab:numbers} for the interaction types in \Tab~\ref{tab:simmodela}, for the resonances in model~B in \Tab~\ref{tab:simmodelb}, for multi-pion production $M_{p=p'}=0.69$ and $M_{p \neq p'}=0.31$, and for direct production in \Tab~\ref{tab:numbers} for the interaction types in \Tab~\ref{tab:simdirectmodel}. The inelasticities can in resonance model~A be obtained from the $\chi^{\mathrm{IT}}$ in  \Tab~\ref{tab:simmodela} according to \equ{kchi}, \ie, $K^{\mathrm{IT \, 1,2,3}} = \sum_\pi \chi_{p \rightarrow \pi}^{\mathrm{IT}} \, M^{\mathrm{IT}}_\pi$, where the number of pions produced $M^{\mathrm{IT}}_\pi=1$ for ITs~1, 2a, and 2b, and $M^{\mathrm{IT}}_\pi=2$ for IT~3 (for IT~2, the pions in 2a and 2b are summed over). For resonance model~B, they are given in
\Tab~\ref{tab:simmodelb}, for multi-pion production $K^{\mathrm{Multi-\pi}} \simeq 0.6$, and for direct production
they are listed in \Tab~\ref{tab:simdirectmodel} (here IT $\mathrm{T_{2b}}$ is not counted separately).

The re-injection rate $p \rightarrow p'$ for initial nucleons $p$ can be obtained analogously to \equ{split} and \equ{prodsimple} from
\begin{equation}
R^{\mathrm{IT}}(x,y)= \delta(x - (1-K^{\mathrm{IT}})) \,  M_{p'}^{\mathrm{IT}} \, f^{\mathrm{IT}}(y)
\label{equ:splitreinj}
\end{equation}
as
\begin{equation}
Q_{p'}^{\mathrm{IT}} = N_p \left( \frac{E_{p'}}{1-K^{\mathrm{IT}}} \right) \frac{m_p}{E_{p'}} \int\limits_{\epsilon_{\mathrm{th}}/2}^\infty dy \, n_\gamma \left( \frac{m_p \, y \, (1-K^{\mathrm{IT}})}{E_{p'}} \right) \,  M_{p'}^{\mathrm{IT}}  \, f^{\mathrm{IT}}(y) \, .
\label{equ:reinj}
\end{equation}
Note that double counting of the same interaction has to be avoided. In particular, interaction type~2 must not be counted twice.\footnote{Although there are two pions produced, there is only one secondary nucleon. For the inelasticity, however, the energy losses into all pions have to be taken into account. For the interaction rate, ITs 2a and 2b are counted as one.} In addition, note that for multi-pion production \equ{mp} based on the total cross section should be used in all cooling, escape, and re-injection rates.

\section{Benchmarks}
\label{app:benchmarks}

Our benchmarks are given in the SRF. The benchmark for GRBs taken from  \citet{Lipari:2007su}. The photon spectrum, a broken power law, is given by
\begin{equation}
 n_\gamma(\epsilon)=\begin{cases}
                     \mathrm{N}_\gamma \, \left(\frac{\varepsilon}{\mathrm{GeV}}\right)^{-1} \frac{1}{\mathrm{GeV \, cm^3}}&  0.2 \,\mathrm{eV} \leq \varepsilon \leq 1 \, \mathrm{keV}\\
		     \mathrm{N}_\gamma\, 10^{-6} \left(\frac{\varepsilon}{\mathrm{GeV}}\right)^{-2} \frac{1}{\mathrm{GeV \,  cm^3}}& 1 \,\mathrm{keV} \leq \varepsilon \leq 300 \, \mathrm{keV}
                    \end{cases}
\end{equation}
and the proton spectrum by
\begin{equation}
 n_p(E_p)= \mathrm{N_p} \,  \left(\frac{E_p}{\mathrm{GeV}}\right)^{-2} \exp{\left[-\left(\frac{E_p}{6.9\cdot10^8\,\mathrm{GeV}}\right)^2\right]}\frac{1}{\mathrm{GeV \, cm^3}} \qquad E_p \geq 1\, \mathrm{GeV} .
\end{equation}
Note that there are dimensionless normalization constants N$_\gamma$ and $\mathrm{N_p}$. The resulting neutrino spectrum is characterized by a wide maximum in $E^2 Q_\nu(E)$.
This benchmark is designed to fit an average of the total distribution of GRB. We limit ourselves to this average, since taking all extreme cases of GRB would go beyond the scope of this paper.

The benchmark for AGNs is adopted from \citet{Mucke:2000rn}. The photon spectrum, a power law with a sharp cutoff, is given by
\begin{equation}
 n_\gamma(\epsilon)=\begin{cases}
                     \mathrm{N}_\gamma \, \left(\frac{\varepsilon}{\mathrm{GeV}}\right)^{-1.6}\frac{1}{\mathrm{GeV \, cm^3}} & 10^{-3}\,\mathrm{eV} \leq \varepsilon \leq 140 \, \mathrm{eV}\\
		    \mathrm{N}_\gamma \, \left(1.4 \cdot 10^{-7} \right)^{0.2}\,\left(\frac{\varepsilon}{\mathrm{GeV}}\right)^{-1.8}\frac{1}{\mathrm{GeV \, cm^3}} & 140 \,\mathrm{eV} \leq \varepsilon \leq 3.6 \, \mathrm{keV}
                    \end{cases}
\end{equation}
and the proton spectrum is given by
\begin{equation}
 n_p(E_p)=\mathrm{N_p} \, \left(\frac{E_p}{\mathrm{GeV}}\right)^{-2} \exp{\left[-\left(\frac{E_p}{2.6\cdot10^9 \,  m_p}\right)^2\right]} \frac{1}{\mathrm{GeV \, cm^3}} \qquad E_p \geq 1\,\mathrm{GeV}
\end{equation}
with $m_p=0.938\,\mathrm{GeV}$. This benchmark is well in the range of usual parameters of the HBL (a subclass of AGN), which are the most interesting objects for state-of-the art Air Cerenkov Telescopes.

The third benchmark, the most challenging one, is a high energetic black body spectrum (BB), adopted from \citet{Mucke:1998mk}. The photon spectrum of temperature $10\,\mathrm{eV}$ is given by
\begin{equation}
 n_\gamma(\epsilon)=
                     \mathrm{N}_\gamma \, 1.318\cdot10^{31}\left(\frac{\varepsilon}{\mathrm{GeV}}\right)^{2}\frac{1}{\exp\left[\frac{\varepsilon}{\mathrm{GeV}}\cdot10^8\right]-1}\frac{1}{\mathrm{GeV \, cm^3}}
\end{equation}
and the proton spectrum with a sharp cutoff is
\begin{equation}
 n_p(E_p)=\begin{cases}
                     \mathrm{N}_p \, \left(\frac{E_p}{\mathrm{GeV}}\right)^{-2}\frac{1}{\mathrm{GeV \, cm^3}} & 10^{6}\,\mathrm{GeV} \leq E_p \leq 10^{12} \, \mathrm{GeV}\\
		    0 & \mathrm{else.}                    \end{cases}
\end{equation}
The blackbody temperature is designed to fit usual BLR photons.
It is the most challenging benchmark because the proton spectrum is not smeared out for the highest energies by an exponential cutoff and the photohadronic interactions are dominated by high energetic multi-pion production due to the considered peaked photon spectrum. 
Even though the benchmark is adopted from \citet{Mucke:1998mk}, it does not exactly represent real physics. Thermal spectra are usual produced outside the shock and are therefore beamed. Only for the production of cosmogenic neutrinos with CMB photons beaming is negligible \citep{1979ApJ...228..919S}

\section{Comparison with SOPHIA runs}
\label{app:SOPHIA}

In SOPHIA, an injected proton of energy $E_p$ is assumed to interact for sure, and the secondary particle (of type $b$) distribution $d n_{p \rightarrow b}/dE_b (E_p,E_b)$ is computed for a specific proton energy or a range of proton energies. This means that
\begin{equation}
 \int \frac{d n_{p \rightarrow b}}{dE_b} (E_p,E_b) dE_b = N_b
\label{equ:nb}
\end{equation}
is the number of secondary particles produced.
The particle spectrum can then be computed using
\begin{equation}
Q_b(E_b) = \int dE_p \, N_p(E_p) \, \Gamma_{p\rightarrow b} (E_p) \, \frac{d n_{p \rightarrow b}}{dE_b} (E_p,E_b) \, .
\label{equ:prodsophia}
\end{equation}
Note that this formula is very similar to \equ{prod}, but not split up in different interaction types.

As it is obvious from \equ{prodsophia}, the interaction rate as a function of proton energy is needed as additional input. We use \equ{pgamma2} with the total cross section as depicted in \figu{xsec} and parameterized in \Sec~\ref{sec:xsec} to compute the interaction rate (the cross section is already summed over all interaction types then). As far as the units and normalization are concerned, it is useful to specify all energies in GeV and all cross sections in $\mu \mathrm{barn} = 10^{-30} \, \mathrm{cm}^2$ (note, however, that in SOPHIA, photon spectra are always given in eV). In this case, the interaction rate carries units of $3 \cdot 10^{-20} \, \mathrm{N_\gamma} \, \mathrm{s^{-1}}$, where $\mathrm{N_\gamma}$ is the dimensionless normalization of the photon spectrum in \App~\ref{app:benchmarks}. From \equ{prodsophia}, it is obvious that $Q_b$ comes in units of $3 \cdot 10^{-20} \, \mathrm{N_\gamma} \, \mathrm{N_p} \, \mathrm{s^{-1}} \, \mathrm{GeV^{-1}} \, \mathrm{cm^{-3}}$ if $N_p$ is given in units of $\mathrm{GeV^{-1}} \, \mathrm{cm^{-3}}$ and carries the (dimensionless) normalization factor $\mathrm{N_p}$. The same units apply to the results from our simplified models.

SOPHIA uses logarithmic energy spacing in $E_p$ and $E_b$ and provides the output on a discretized energy grid in the form $d n_{p \rightarrow b}/dx_b (E_p,x_b)$ with $x_b= \mathrm{log} ( E_b/E_p)$, equally spaced in
$\Delta x_p$ with $x_p= \mathrm{log} ( E_p/ \mathrm{GeV})$ and in $\Delta x_b$. For the easiest data extraction, it is advisable to use $\Delta x_p=\Delta x_b$.
Then, it is useful to re-write \equ{prodsophia} as
\begin{equation}
Q_b(x_b^j) = \Delta x_p \sum_{i}  \, N_p(x_p^i) \, \Gamma_{p\rightarrow b} (x_p^i) \, \frac{d n_{p \rightarrow b}}{dx_b} (x_p^i,x_b^j) \, 10^{-x_b^j} \, ,
\label{equ:sdis}
\end{equation}
where the last factor comes from switching to the logarithmic scale. With \equ{sdis}, the SOPHIA output can be used directly, collecting all entries for a specific $x_b^j$ from all proton energy bins.  Note, however, that still the particular output format has to be taken into account (SOPHIA first lists the bin range filled with data, then the filled bins, as a function of the proton energy).

For the test runs of the AGN and GRB spectra, we use a proton energy grid between $1 \, \mathrm{GeV}$ and $10^{10} \, \mathrm{GeV}$ with 100 bins, \ie, $\Delta x_p=0.1$. In addition, we use 100 output bins with a step size $\Delta x_b=0.1$. For the GRB benchmark, we use 100000 trials per proton bin, for the AGN benchmark 25000 trials. For the BB test run, we use a proton energy grid between $10^6 \, \mathrm{GeV}$ and $10^{12} \, \mathrm{GeV}$ with 60 bins, \ie, $\Delta x_p=0.1$. In addition, we use 75 output bins with a step size $\Delta x_b=0.1$ and 10000 trials per proton bin.

\end{appendix}

\end{document}